\documentclass[acmsmall]{acmart}

\AtBeginDocument{%
  }

\setcopyright{acmlicensed}
\copyrightyear{2024}
\acmYear{2024}
\acmDOI{XXXXXXX.XXXXXXX}

\acmJournal{TOCHI}
\acmVolume{0}
\acmNumber{0}
\acmArticle{0}
\acmMonth{0}

\begin{document}

\title{Towards improved software visualisation of parameterised REE patterns: Introducing REEkit for geological analysis}

\author{Jaxon Kneipp}
\email{jaxon.kneipp@gmail.com}
\author{Alex Potanin}
\email{alex.potanin@anu.edu.au}
\author{Michael Anenburg}
\email{Michael.Anenburg@anu.edu.au}
\affiliation{%
  \institution{Australian National University}
  \city{Canberra}
  \state{ACT}
  \country{Australia}
}

\renewcommand{\shortauthors}{Trovato et al.}

\begin{abstract}
Modern geological studies and mineral exploration techniques rely heavily on being able to digitally visualise and interpret data. Advanced technologies like portable X-Ray fluorescence (pXRF) have transformed how we collect and analyse geochemical data, including valuable information about rare earth elements (REEs). REEs are vital for renewable energy technologies, with global demand expected to rise significantly. REE concentrations, when normalised to a standard material, show unique geometric curves (or patterns) in geological samples due to their similar chemical properties. The lambda technique can be used to describe these patterns and turn them into points - making it easier to visualise an interpret larger datasets. Lambdas have the potential to help industry understand intricate sample relationships and the geological and economic importance of their data. Lambda data can also greatly enhance the utility and value of large datasets in the mineral exploration space.

This study explored the use of lambdas through the evaluation of various visualisation methods to determine their usefulness in mineral exploration. Additionally, the study identified the visualisations that are effective at enhancing lambda data both independently and in combination. This was achieved through the development of the web application ‘REEkit’. This platform facilitated the evaluation of the different visualisation methods and gauged industry interest and acceptance of such a service. Qualitative data was gathered through contextual inquiry, utilising semi-structured interviews and an observational session with 10 participants. Conceptual thematic analysis was applied to extract key findings and conclusions from this research.

This study found that two critical factors for successful lambda data visualisation in the mineral exploration industry are familiarity and clarity, particularly when compared to existing REE visualisation methods such as spider diagrams. Meaning that visualisations that were familiar and common place for users allowed for better analysis and clear communication to non-technical audiences when using lambda data. This included visualisations such as the 3D scatter plot and scatter plot matrix. Furthermore, visualisations that complemented each other and seamlessly integrated into the same workflow provided diverse perspectives on the data. Important aspects included understanding population grouping versus data distribution, achieved through combinations such as scatter plot and density contour plot, or 3D scatter plot and violin plot. These contributions provide a basis for future research in this area, indicating that visualisation choices are strongly shaped by individual workflows and goals.
\end{abstract}

\maketitle

\section{Introduction}

Using illustrations and visualisations to describe the natural world around us has formed an integral part of modern geological research. In recent years, technological advancements have changed the way we collect, process, and analyse geological data; more specifically, geochemical data or data pertaining to the chemistry and composition of geological samples. Traditional field sampling and lab analysis methods have been improved, and in some cases replaced, by automated processes and advanced field instruments such as the portable X-Ray fluorescence (pXRF) machine, a handheld device using X-ray fluorescence for non-invasive elemental analysis of samples. 

This gives geologists access to data and insights which were previously unattainable on this scale. An example of such data that has become increasingly more accessible and advantageous to collect is rare earth element (REE) data. 

Comprising of 14-16 elements (the Lanthanide series – from lanthanum to lutetium, excluding promethium and sometimes including scandium and yttrium) these metals are an essential component of electric vehicles (EVs), wind turbines and other technologies essential for the global transition to renewable energy due to their magnetic and catalytic properties. As a result, it has been speculated that demand for the REEs could increase by up to 70\% in the next decade (Adams, 2022).

REEs are often found together in geological samples and deposits due to their similar chemical behaviour (e.g., similar ionic charge and radii). However, the slight changes in REE chemical behaviour affect their partitioning during geological processes such as weathering, fractionation (magma evolution) and mineral formation (Dushyantha et al., 2020). When depicted graphically, the relative concentration of each REE (normalised to a reference standard, typically the average composition of the solar system) form a smooth pattern which can be compared between samples. An example of these curves is shown in Figure 1.

\begin{figure}
\includegraphics[width=14cm]{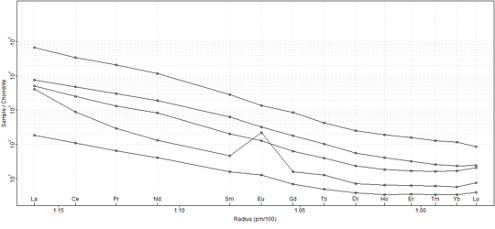}
\caption{Figure 1: A spider diagram depicting a series of rare earth element patterns. Elements are represented on the x-axis by atomic radius in picometers, while the sample’s normalised concentration is plotted on the y-axis.}
\end{figure}

The shape of these patterns can give insight into the geological formation of samples in addition to the economic implications for the region it was taken from. For example, a U-shaped curve would be enriched in both light and heavy rare earths, suggesting potential fluid alterations, while a flat pattern would suggest that the sample was taken from a homogenous magma source. Comparing REE bearing samples in this way is common practice both in academia and the mineral resources industry. However, visualisation and interpretation of large pattern datasets can produce dense and cluttered plots which are hard to interpret and often yield little insight.

In recent years, there has been growing research into refining and parameterising REE patterns, with the objective of enhancing the value of larger REE datasets through providing improved visualisations and illustrations to academics and industry researchers. Work done at the Australian National University (ANU) has led to the development of a technique which can be used to help describe the shape of each REE pattern, through constants known as ‘lambda’ values. Throughout this thesis, we refer to this technique as the lambda technique/method. The most commonly used lambda values include:

\begin{itemize}
\item $\lambda$0 : Describes the abundance of the REEs within a sample (magnitude of the pattern)
\item $\lambda$1 : Describes the linear slope of the pattern
\item $\lambda$2 : Describes the quadratic curvature of the pattern
\item $\lambda$3 : Describes the degree of inflection at the ends of the pattern (sinusoidality)
\end{itemize}

Examples of lambda values for a series of three different curves can be seen in Figure 2.

\begin{figure}
\includegraphics[width=14cm]{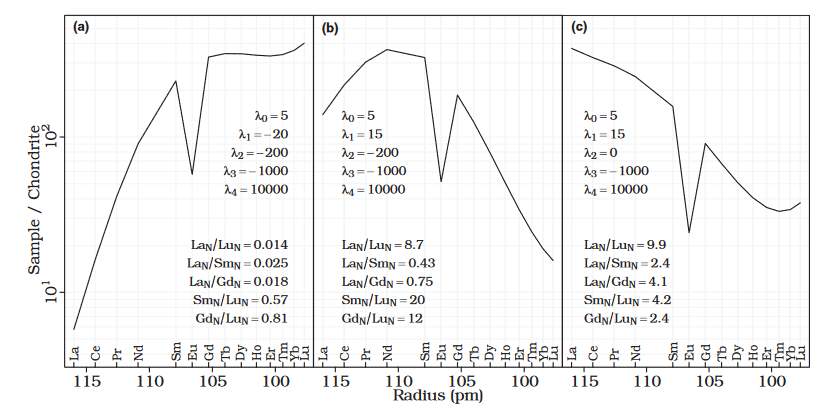}
\caption{Figure 2: Three unique rare earth element patterns (a, b, and c) shown with corresponding lambda values ($\lambda$) and elemental ratios. Elements are represented on the x-axis by atomic radius in picometers, while sample concentration is plotted on the y-axis. Image source: (Anenburg, 2020).}
\end{figure}

This project seeks to extend this field of research by exploring six common visualisation methods (\textbf{spider diagram, scatter plot, 3D scatter plot, scatter plot matrix, density contour plot and violin plot}) to find which provides better insight for a given collection of REE patterns and their associated lambda data. Our research will achieve this goal while also overseeing the development of a software visualisation tool (REEkit) which can be used by both academics and industry to utilise REE lambdas in their research. 

The evaluation of the six visualisation methods will be done through contextual inquiry, a method aimed at addressing the following research questions:

\begin{itemize}
\item In what specific contexts does lambda data provide enhanced insights?
\item Among the visualisation methods, which ones best present lambda data when viewed in isolation?
\item How can different visualisation methods be integrated to provide deeper geological insights using lambda data?
\end{itemize}

Through contextual inquiry, these questions will be investigated to gain better insight into the effectiveness and applicability of different visualisation methods in the context of lambda data.

\section{Background}

In this literature review, we aim to explore the existing body of literature pertaining to the visualisation of REE data. Particular emphasis will be put on the quantification of REEs and current metrics for interpreting REE patterns, visualisation of REE geochemical data, and evaluating software tools currently accessible for such visualisation purposes. Additionally, a review of software visualisation evaluation methods, specifically contextual inquiry (the primary method of data collection for this study), will also be presented. We conducted this review by examining publicly available academic documents (such as conference and journal papers). This is an active field of research and as such it is important to note that there may be unpublished work in this space which is omitted from this review. 

\subsection{Advancements in reporting methods for REE patterns}

\subsubsection{Quantification of REEs: current metrics and approaches}

REE composition in geological samples is important both from an economic and a geochemical perspective. Rare earth elements serve as a valuable tool in geochemistry, as they exist as trace elements in a wide range of natural geological samples collected worldwide. A trace element is defined as an element whose concentration comprises less than 0.1\% of a rock's composition (Kennedy, 1998). Analysing trace element concentrations can offer valuable insights for geochemical research, leading to a high demand for data on REE concentrations even prior to the recent economic interest.

Currently, exploration companies employ a reporting metric called total rare earth oxides (TREO) for quantitative reporting of REEs (an example of this can be seen in Figure 3). However, relying solely on TREO as a metric to evaluate the economic viability of a potential deposit has inherent limitations. To illustrate this point, let's consider a sample with a relatively high TREO percentage of 10\%. Upon closer examination, it becomes evident that the majority of the oxides present are lanthanum and cerium. In 2022, the market prices for these metals were approximately \$2 and \$1.50 USD per kilogram of oxide respectively (Gielen and Lyons, 2022). On the other hand, another deposit may exhibit a lower TREO percentage of 1\%, but it contains significant quantities of neodymium and praseodymium, each valued at around \$140 USD per kilogram (Gielen and Lyons, 2022). In this scenario, despite having a lower TREO value, the second deposit proves to be significantly more economically valuable than the first. This example clearly demonstrates how relying solely on TREO (as many companies do) can be misleading when evaluating drilling results, as it fails to account for variations in individual rare earth element prices and their relative economic significance. 

\begin{figure}
\includegraphics[width=8cm]{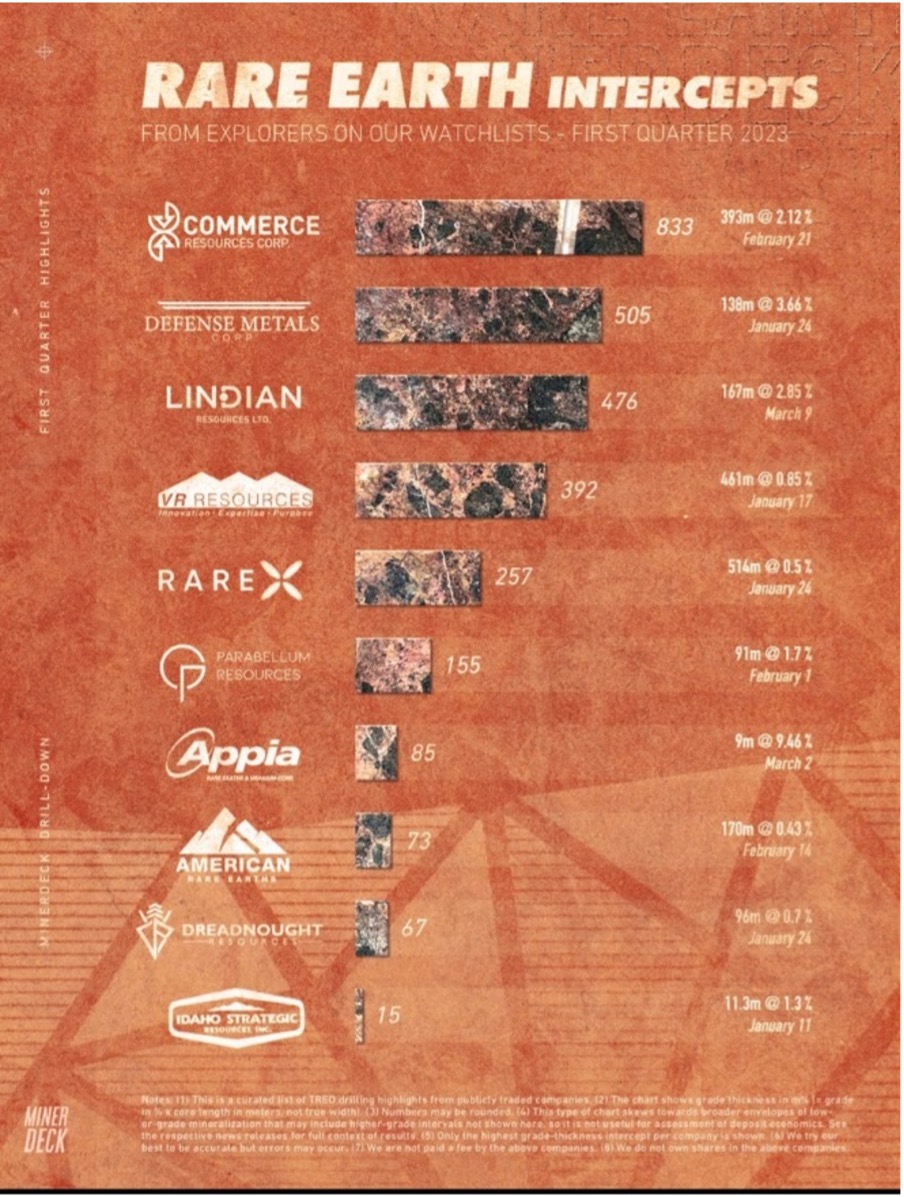}
\caption{Figure 3: Promotional material published on LinkedIn by MineDeck highlighting the best drilling results for the first quarter of 2023. These companies are ranked based on the TREO percentage (seen at the end of each row, surrounded by the black box) of their intercepts (MinerDeck, 2023).}
\end{figure}

Alternative approaches to reporting REEs involve the use of elemental ratios, such as the neodymium and praseodymium (NdPr) ratio (which is the proportion of Nd and Pr relative to other REEs) or the ratio of light rare earth elements (LREE) to heavy rare earth elements (HREE) (LREE/HREE). These methods offer a more insightful perspective on the economic value of deposits. However, they still have limitations that become evident when considering the entire suite of REEs. This is because ratios only look at small parts of the REE suite relative to each other and do not consider the entire REE series as a whole. These approaches provide valuable information but leave certain gaps in the overall assessment, which could be addressed by utilising a reporting metric that considers the complete set of REEs. The lambda method addresses this by introducing a set of metrics known as lambda values. These values facilitate a more comprehensive reporting metric for drilling results by encompassing the entire series of REEs. 

By utilising the lambda method, investors, stakeholders and geoscientists can gain a better and standardised understanding of the complete picture, rather than being limited to a narrow perspective that a company may selectively present. This enhanced reporting approach empowers investors to make more informed decisions by considering the whole series of REEs when reading and analysing reported results. 

\subsubsection{Advancements in the lambda method for REEs}

Distilling information from REE pattern curves has been explored by several ANU researchers in recent years. Hugh O’Neill pioneered this through his research with basaltic rock samples in 2016. The study by O’Neill (2016) discusses how precise quantification of REE patterns (treated as mathematical objects) would facilitate being able to more easily compare larger datasets of REE patterns from samples. O'Neill's paper provides a comprehensive explanation of the underlying theory behind the quantification approach, which involves a detailed outline of the steps required to derive lambda values from REE patterns; essentially fitting a polynomial function to the curve (through the least squares method) and extracting the functions coefficients to get lambda values. Accompanying these steps, the study provides an excel spreadsheet as supplementary material which allows researchers to experiment with the method on their own data.

O’Neill also discusses the potential pitfalls and limitations that may arise when employing this method. One notable concern highlighted is the possibility of encountering misfits and less accurate lambda values due to certain aspects of curves, such as elemental anomalies. After applying the approach to a series of patterns from basaltic samples, O'Neill highlights the significant impact this technique could have on ongoing petrogenetic and geochemical studies. While he did not specifically address the potential implications for suites of rocks containing economic levels of REEs, his work still provides valuable insights for researchers looking to utilise this approach for quantification purposes.

Expanding upon O'Neill's research, a subsequent paper by Anenburg (2020) examined the method at a broader level, delving into the influence of anomalous elements (e.g. cerium and europium) on the accuracy of the lambda method. This study explored how lambdas could be utilised to model the speciation of REE minerals in the Earth's crust, enabling the prediction of potential mineral assemblages that have not yet been discovered but are likely to exist. The study not only highlighted the diverse applications of O'Neill's method in the field of geochemistry but also improved the method's accessibility by presenting a user-friendly web app built in R. This software tool is further discussed in section 2.3. The study by Anenburg (2020) expands on some other documented metrics for describing REE pattern shape such as ratios, i.e., $\frac{La}{Lu}$ , $\frac{La}{Gd}$, and $\frac{Gd}{Lu}$ (these can also be seen in Figure 2 for a series of curves). 

Anenburg highlights a crucial point: ratios alone cannot fully capture the complexities inherent in certain REE patterns, such as curvature or sinusoidality. This observation sheds additional light on why the lambda data is preferable when working with REE patterns. By deriving lambda values, a more comprehensive understanding of the intricate features present in REE patterns is offered. This insight, which O'Neill did not explicitly address, further emphasises the advantages and relevance of the lambda method in analysing and interpreting REE patterns.

In Anenberg’s study, a notable limitation identified was the lambda method's inability to capture pattern variability caused by the tetrad effect. The tetrad effect is a chemical phenomenon which affects several REEs. It is related to the filling of the 4f electron shell (McLennan, 1994). The joint study by (Anenburg and Williams, 2022) investigated this limitation more comprehensively, outlining additional analytical techniques to quantify the impact of the tetrad effect, resulting in the generation of "tetrad coefficients" for REE patterns. By addressing this limitation and introducing additional analytical techniques, the study improved the lambda method's capabilities and provides researchers and geoscientists with valuable insights for better understanding and accounting for the tetrad effect in their analyses.

The development of the lambda method has followed a sequential path, with various authors agreeing on the advantages it brings to the field of geochemistry. Over time, incremental advancements have led to the refinement of the method we have today. These improvements have contributed to enhancing its accuracy, applicability, and overall effectiveness. 

O'Neill's speculation about the potential of the method to enable easier REE pattern comparison raises an important question: how can this be achieved? This is where our study comes into the picture. The objective of our study is to investigate the optimal visualisation of the end result of the method (the lambda values themselves) to ensure meaningful insights and effective comparison between geological samples and their REE patterns. 

\subsection{Exploring techniques for REE and geochemical visualisation}

The visualisation of REE data has a longstanding history, dating back to the 1960s as documented in the study by Coryell, Chase and Winchester (1963). This seminal study introduced the process of normalisation and visualisation of REE patterns. The authors recognised the challenges associated with interpreting non-normalised "zig-zag" REE data (Figure 4), which can be attributed to irregularities in REE abundances on earth resulting from varying levels of cosmic production.

Consequently, the study emphasised the necessity of normalising REE concentrations to chondrite concentrations, which provide the best estimate of the Earth's primordial composition. Chondrites are meteorites which provide us with the best estimate of earths primitive composition (Scott and Krot, 2007). By applying this normalisation technique, a much smoother REE curve/pattern can be obtained, making patterns easier to interpret. Traditionally, these patterns have been visualised using spider diagrams (also known as spidergrams). Spidergrams (seen in figure 1 and 4) have become a widely adopted visualisation technique in the field of geochemistry for plotting elemental compositions and at the heart of the lambda method

\begin{figure}
\includegraphics[width=14cm]{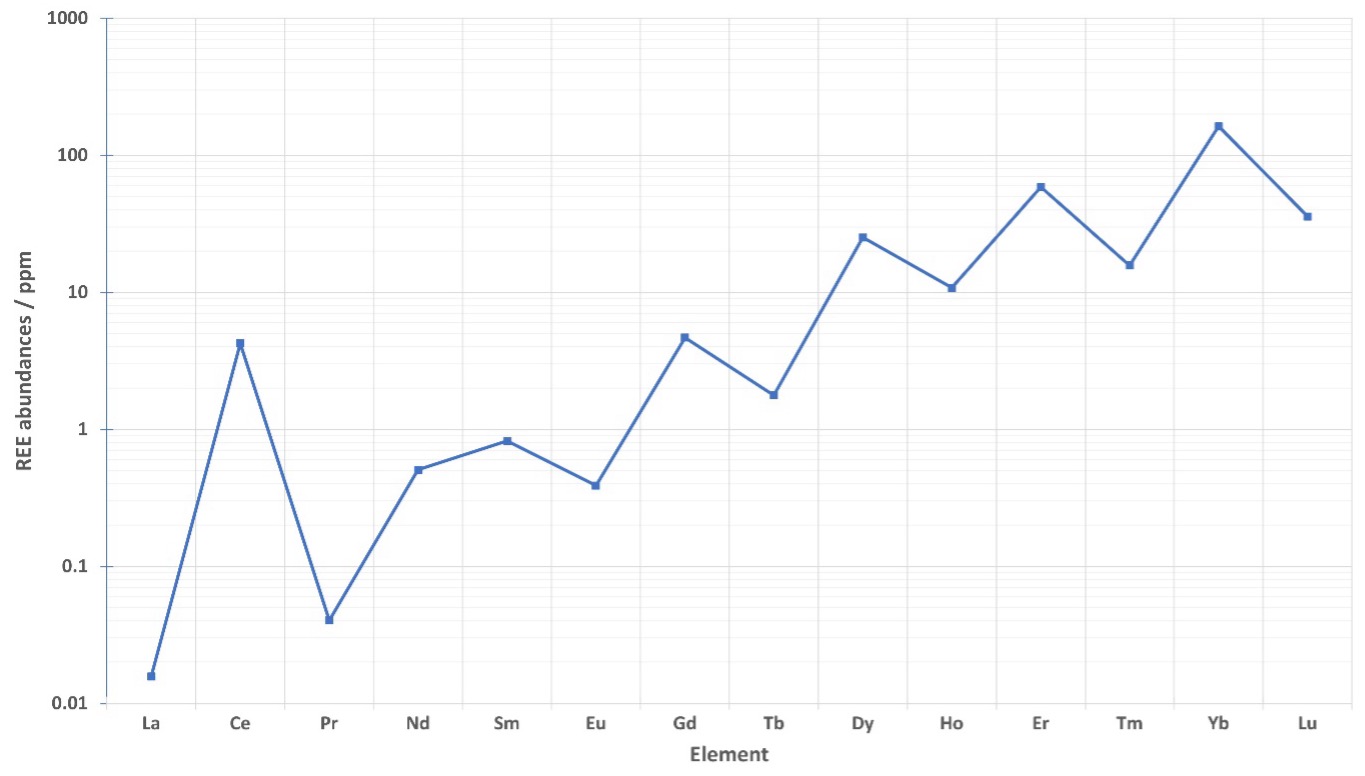}
\caption{Figure 4: A spider diagram depicting a rare earth element pattern. Elements are represented on the x-axis by atomic radius in picometers, while samples concentration is plotted on the y-axis. Note the ‘zig zag’ nature of the pattern.}
\end{figure}

In subsequent studies, researchers have explored alternative methods by employing spidergrams to analyse REE concentrations but normalising values to a range of standards. These standards include the composition of the present-day mantle, mid-oceanic ridge basalt (MORB) and many more. A commentary released in 1987 drew attention to the pervasive miscommunication resulting from the use of different normalisation standards (Rock, 1987). The commentary emphasised the critical need to establish standardised normalisation values.

Within the commentary, Rock presented an argument advocating for a reduction in the pool of standards to just chondrite and MORB (although others including PAS, for mafic rocks, and NASC, for crustal rocks, are also used today), asserting that such an approach would be sufficient for the majority of igneous rocks and are the most widely used. This viewpoint has gained widespread acceptance in contemporary research and is commonly adopted by scientists today. By minimising the pool of normalisation standards to chondrite and MORB, researchers have achieved greater consistency and comparability in their findings when studying REE concentrations in igneous rocks using spidergrams.

In their textbook, titled "Using Geochemical Data to Comprehend Geological Processes", authors Hugh Rollinson and Victoria Pease summarise another noteworthy visualisation technique for analysing REE data: ratio diagrams (Rollinson and Pease, 2021). These plots, displaying the ratios of different REEs across a sequence of samples, offer valuable insights into specific petrogenetic and geological scenarios. For instance, they can reveal the extent to which LREE fractionate compared to HREE. Ratio diagrams (like those seen in figure 5) serve as a powerful tool in understanding the intricate relationships and dynamics within nuanced petrogenetic processes. 

\begin{figure}
\includegraphics[width=14cm]{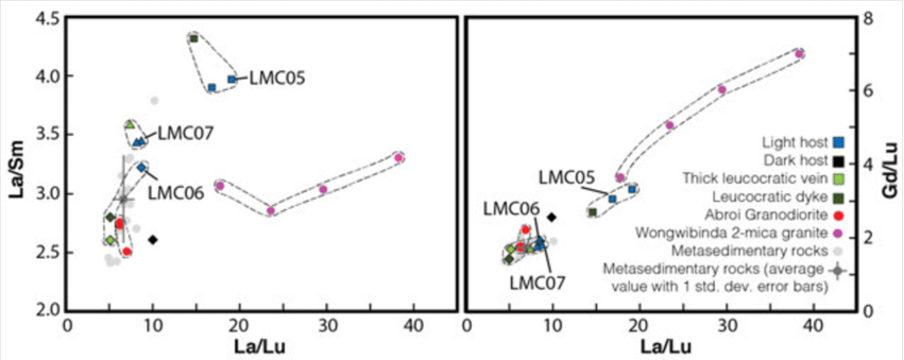}
\caption{Figure 5: An example of a ratio diagram used to describe REE data from a series of rocks taken from Wongwibinda, NSW (Farmer, 2017).}
\end{figure}

The predominant methods of data interpretation in the realm of REEs are centred around these two key visualisation techniques (spider and ratio diagrams). They form the cornerstone of analysing REE data, with slight variations influenced by researchers' individual preferences. As technology has progressed, software visualisation of such data has become increasingly popular in this space with a wide variety of tools flooding the market, each promoting slight differences that will aid researchers in getting better insight from their data (some of these tools are discussed more in section 2.3). 

The lambda method has yet to permeate to the forefront of this field and as such there has been no research done into what visualisation methods best suite the data type. Traditional techniques such as spidergrams and ratio diagrams are rendered ineffective since lambda data no longer conforms to the structure necessary for their application. As a result, there is a critical need to explore different visualisation methods to accommodate the unique characteristics of lambda data.

Some documented methods we plan to employ during this research are standard scatter plots as well as density contour plot, 3D scatter plots and violin plots. We will also investigate what visual aids best suit the outlined methods and data type. The web-based software application (REEkit) will be developed in this study to facilitate this investigation. REEkit will complement existing software tools and aim to streamline the processing, visualisation, and analysis of REE geochemical data, providing a user-friendly experience for researchers and industry professionals. 

\subsection{Survey of REE data visualisation software}

In this section, we'll introduce some key software tools that have advanced the field of geochemical data analysis. By examining these tools, we aim to identify key aspects that we intend to incorporate into the development of REEkit. Additionally, we will discuss the limitations associated with the current tools, shedding light on the areas where further improvement is necessary. This analysis will serve as a foundation for understanding the existing landscape and informing the development of REEkit, ensuring that it addresses the needs and challenges faced by geoscientists and industry professionals working with REE data.

Software tools have been developed to assist geoscientists over the last 35 years, having to meet the demands of larger and increasingly more complex datasets as the years go by (Godwin, Valleau and Mortimer, 2021). Over the past decade, there has been a noticeable shift among researchers, transitioning away from relying on generic software tools like Excel and Numbers. Instead, there is a growing preference for custom software developed in data science-friendly languages such as R and Python. This transition is driven by the increasing appeal of bespoke software that caters specifically to the needs of scientists. By leveraging the capabilities of R and Python, scientists can harness advanced data analysis and visualisation techniques, enabling more efficient and insightful exploration of their datasets.

Developed in 2005, “Petrograph” (Petrelli et al., 2005) was an application built to run on Windows 98/2000/XP. It allowed geoscientists to load geochemical data into the program in widely used formats such as Excel documents. Once data was loaded in users would be able to investigate it using established geochemical visualisation techniques such as ternary and classification diagrams. Petrograph enabled users to create trace element spidergrams, yet its functionality regarding REE was limited. It was confined to generating these spidergrams and did not encompass broader REE capabilities within the platform. Despite this, the application was revolutionary at the time as it was able to handle larger datasets without the need to transfer data between different pieces of software – it was the ‘all-in-one’ package. While Petrograph is all but obsolete today, it set the tone for future applications and packages involving geochemical data. 

More recently, “GeoPyTool” (Yu et al., 2019) was released. The paper by Yu et al. describes the tool as a means of converting raw geochemical data from Excel files into rich graphical visualisation files. It boasts a set of functionalities which supersede that of Petrograph and runs on most popular operating systems (Windows, Mac, and Linux). The interface of GeoPyTool provides an enhanced user experience, aligning with modern desktop standards. The tool offers improved visual aids, particularly for spidergrams, further enriching the analysis process. However, most of the functional advancements primarily focus on general geochemical visualisations, with only minor improvements specifically related to trace and REE data visualisation. In addition, it does not take advantage of the lambda method in its calculations and visualisations.

A year following the release of GeoPyTool, “Pyrolite” (Williams et al., 2020) was released. Unlike other software, this tool did not have an accompanying user interface – instead it served as a python package which researchers could utilise while developing their own Python scripts to analyse their data. The modularity of the pyrolite package sets it apart from other software in this field. The paper by Williams (2020) discusses the challenges with dealing with compositional data within the geoscience community and how Pyrolite seeks to simplify the process of analysis. Notably, Pyrolite incorporates the use of density spider diagrams for analysing large collections of REE patterns (see Figure 6). It is important to note that while density spider diagrams offer a simplified representation, they lack the level of detail that would be captured by a tool that leverages the lambda method.

\begin{figure}
\includegraphics[width=14cm]{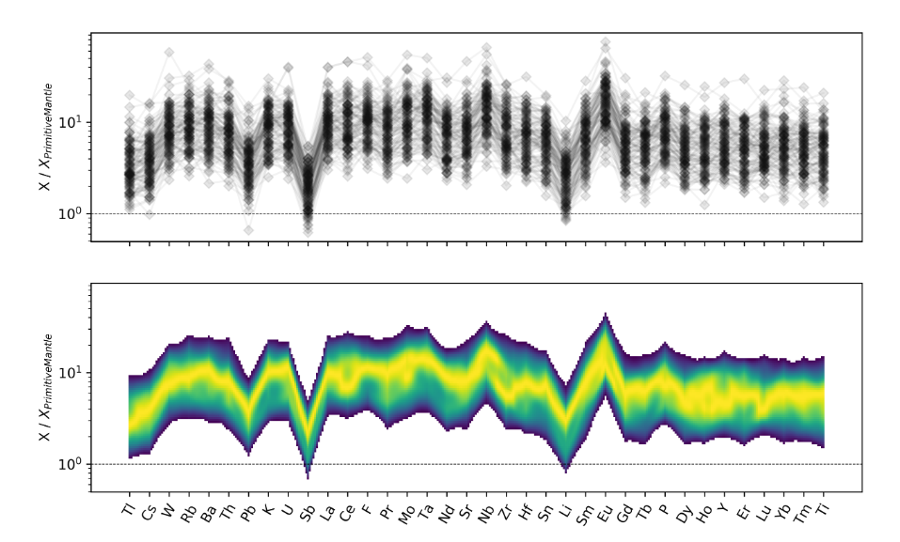}
\caption{Figure 6: An example of a density spider diagram (bottom) compares to a conventional spider diagram (top) (Williams et al., 2020)}
\end{figure}

Two additional tools are also worth noting – GCDkit and Igpet. GCDkit, built in R, offers a user interface and specialised tools for handling whole-rock major- and trace-element analyses as well as isotopic data analysis from igneous rocks (Erban et al., 2003). It supports the plotting of various types of geochemical data including REE data in the form of spidergrams. The tool faces challenges however related to version compatibility, impacting its adoption among users in recent years.

Additionally, Igpet is an open-source piece of software designed for research in igneous geochemistry. It includes tools for rock identification, tectonic discrimination, and various types of numerical modelling (Carr and Gazel, 2017). The tool again, primarily only supports the visualisation of REE data through spidergrams.

While the previously mentioned tools exhibit substantial capabilities in managing geochemical information, they lack specialisation in the utilisation of REE data. This may be attributed to the fact that REE data typically did not require dedicated tools for its analysis, as it primarily involved simply displaying patterns or curves through spidergrams. However, Anenburg addressed this need by developing two web-based software tools, “ALambdaR” and “BLambdaR” (Anenburg, 2020). These tools enable users to apply the lambda method to their REE patterns through a user-friendly web interface. Specifically, ALambdaR and BLambdaR facilitate the conversion of REE data stored in CSV files into lambda values, offering a streamlined workflow for REE analysis. Built using the R programming language, the tools developed by Anenburg offer users the ability to perform basic visualisations, enabling the examination of economic value associated with specific REE patterns. However, it is limited by its inability to compare one pattern to another. 

REEkit aims to enhance and expand upon the functionality of ALambdaR and BLambdaR by providing a specialised interface for sample comparison. This software will be developed with a focus on facilitating effective comparison of REE data across multiple samples using the lambda method. Drawing inspiration from Petrograph, GeoPyTool, and Pyrolite, REEkit will incorporate their best practices in terms of reproducibility, modularity, and handling larger datasets. By combining these beneficial elements, REEkit will provide scientists with a powerful tool for analysing REE data, enabling seamless sample comparison and enhancing the overall workflow for REE analysis.

\subsection{Methods for evaluating information and software visualisation}

Information and software visualisations can be assessed in a wide variety of ways. Information visualisation (Info Vis) is an important field within computer science because it helps humans understand how to represent complex data visually, making it easier to comprehend and manage. Info Vis is highly coupled with human perception, human behaviour and human interaction, and as such it is important that the evaluation methods used in research be human-centered (Nazemi et al., 2015). For this study, qualitative data collection methods are most suitable due to the diverse background of potential users and the study's exploratory nature. Common qualitative methods include contextual inquiry, diary studies, and laboratory observational studies (Carpendale, 2008).

Contextual inquiry is useful for exploratory information visualisation research (such as that being conducted in this study) as it allows researchers to fundamentally understand how users interact with their data in real-life situations. By observing and talking to users directly, researchers are able to learn about their needs and challenges, helping design and implement visualisations that are both insightful and easy to use or adopt. For these reasons, this study has opted for contextual inquiry as its qualitative data collection method (discussed further in section 3.3).

Contextual inquiry has been commonly used in various exploratory information visualisation studies. For instance, Steichen and Fu (2019) investigated the effectiveness of visualisation aids like gridlines and dot grids in helping participants interpret data. They conducted brief interviews with each participant and then had them complete tasks using the talk-aloud protocol, which is the same method employed in this study.

Wang et al. 2019 employed a similar method in their study, in which the research team developed a software platform (DeHumor) to facilitate the interactive analysis of humorous content in speech and voice. The research team assessed its usability and effectiveness through case studies and interviews with expert communication coaches and humour researchers. In this study, we plan to adopt a similar approach through the development of REEkit and subsequent evaluation through interviews and observational studies (contextual inquiry).

This review highlights the relevance of this field in computer science research, especially concerning the visualisation of lambda data. It also highlighted that the method employed in this study has a proven track record, being widely adopted in similar research projects. This validates its applicability and justifies its use in our research.

\subsection{Summary}

As the demand for REEs rises with the shift toward clean energy, the volume of associated data also increases. This makes the need to process, interpret and visualise big data in this space increasingly important. Through this review, we have provided a comprehensive overview of the current state of knowledge in the fields related to the quantification, processing, and visualisation of REE compositional data as well as the current software tools which are used for such tasks.

The current practices for reporting REEs in exploration rely on metrics such as TREO and elemental ratios like NdPr. These metrics have limitations in fully capturing the variation in the suite of REEs, which can pose challenges for both the geoscience community and investors. Section 2.2 of this review investigated the current state of literature concerning the lambda method – a new technique which facilitates better quantification of REE results and allows for potentially improved reporting metrics. 

The method was first devised by O’Neill in 2016 who applied it in a largely geochemical capacity, however it was improved and further utilised by Anenburg and Williams over the past five years. Their contributions saw the implementation of the method in R which could be utilised by users through a web interface (ALambdaR and BLambdaR). During this review we also discussed how our research will fit into the picture by extending the work done by Anenburg (2020) through the development of REEkit which will facilitate the use of the lambda method to visually compare samples on a larger scale.

This review also evaluated the pertinent literature regarding the visualisation of REE data and geochemical data in a broader context. Currently, the spider and ratio diagrams have emerged as the most prominent visualisation methods for analysing REE data, particularly when concentrations are normalised to chondrites. The review also touched on a commentary which advocated for ensuring a small pool of normalisation values – naming chondrites and MORBs as the best standards to use. 

Finally, a survey of the current software tools for conducting the outlined visualisation was also undertaken, highlighting several important features as well as limitations in existing tools. Literature surrounding the release of many of these tools outlines the improvements to user interface, geochemical data processing and visualisation capability in recent years. Petrograph, GeoPyTool and ALambdaR/BLambdaR are three examples of how software tools in this space have evolved since the early 2000s, with the latter being a key inspiration for REEkit. Specifically, Pyrolite is a library which not only highlights the importance of modularity when developing geoscientific software but also illustrates specific advancements in REE visualisation through the density spidergram feature. 

Through this review it has become apparent that the existing literature lacks sufficient support for the lambda method in software solutions. The primary output artifact of this research is to address this deficiency by introducing REEkit, a specialised software tool designed to handle lambda data. Moreover, this study seeks to explore optimal approaches for utilising this data, thereby laying the foundation for future research in this field.

\section{Methodology}

The research strategy and methodology employed in this study involved several distinct stages designed to assess various visualisation methods and enhance the user experience of REEkit.
Initially, informal discussions were held with a small group of potential users, and an extensive literature review was conducted to define the requirements for developing REEkit. 

Following this, the study focused on implementing and evaluating the visualisations presented in REEkit, aligning with the research objectives of understanding when lambda data is useful, finding the best ways to visualise lambda data on its own, and explore how combining visualisation methods can provide deeper geological insights using lambda data.

Data collection primarily took place through contextual inquiry in the form of semi-structured interviews, a widely accepted approach for gathering detailed qualitative data on the effectiveness of different visualisation methods (Carpendale, 2008). Subsequently, transcription and thematic analysis were performed to extract insights and form the study's findings. For a detailed exploration of these results and subsequent discussion, please refer to sections 4 and 5, respectively.

Ethical considerations were also integrated throughout the entire research process, ensuring the responsible conduct of the project in adherence to high ethical standards of the Australian Privacy Act of 1988.

\subsection{Visualisation methods and REEkit development}

\subsubsection{Visualisation techniques}

The visualisation methods evaluated in this study were carefully selected, considering the following factors:

\begin{itemize}
\item \textbf{Multivariate representation} techniques were chosen to effectively show complex relationships in the dataset involving multiple variables (i.e., different lambda values).
\item \textbf{Comparative analysis} techniques help identify patterns, trends, and variations by comparing different sets of data.
\item \textbf{Spatial representation} techniques translate numerical data into unique spatial arrangements.
\item Visualisation techniques tailored to the \textbf{specific needs of geochemistry}, ensuring meaningful portrayal of relevant information.
\end{itemize}

In addition to these factors, the availability of visualisation options in Python played a role in influencing the visualisations selected for this study.
 
This study evaluated the use of the \textbf{spider diagram, scatter plot, 3D scatter plot, scatter plot matrix, density contour plot and violin plot} to visualise lambda data. 

The following sections expand on the specifics of each visualisation method, detailing the rationale behind their inclusion in the study and highlighting the unique features and benefits they offer as a visualisation technique.

\paragraph{Spider diagrams} Spider diagrams, also known as spidergrams (Figure 7), are commonly employed in the field of geochemistry to visualise and analyse trace element concentrations in rocks (Rollinson and Pease, 2021). Each pattern typically represents geochemical data from a specific point in a geological sample. In this study, spider diagrams served as the control, representing the established standard for visualising, and working with REE geochemical data.

\begin{figure}
\includegraphics[width=10cm]{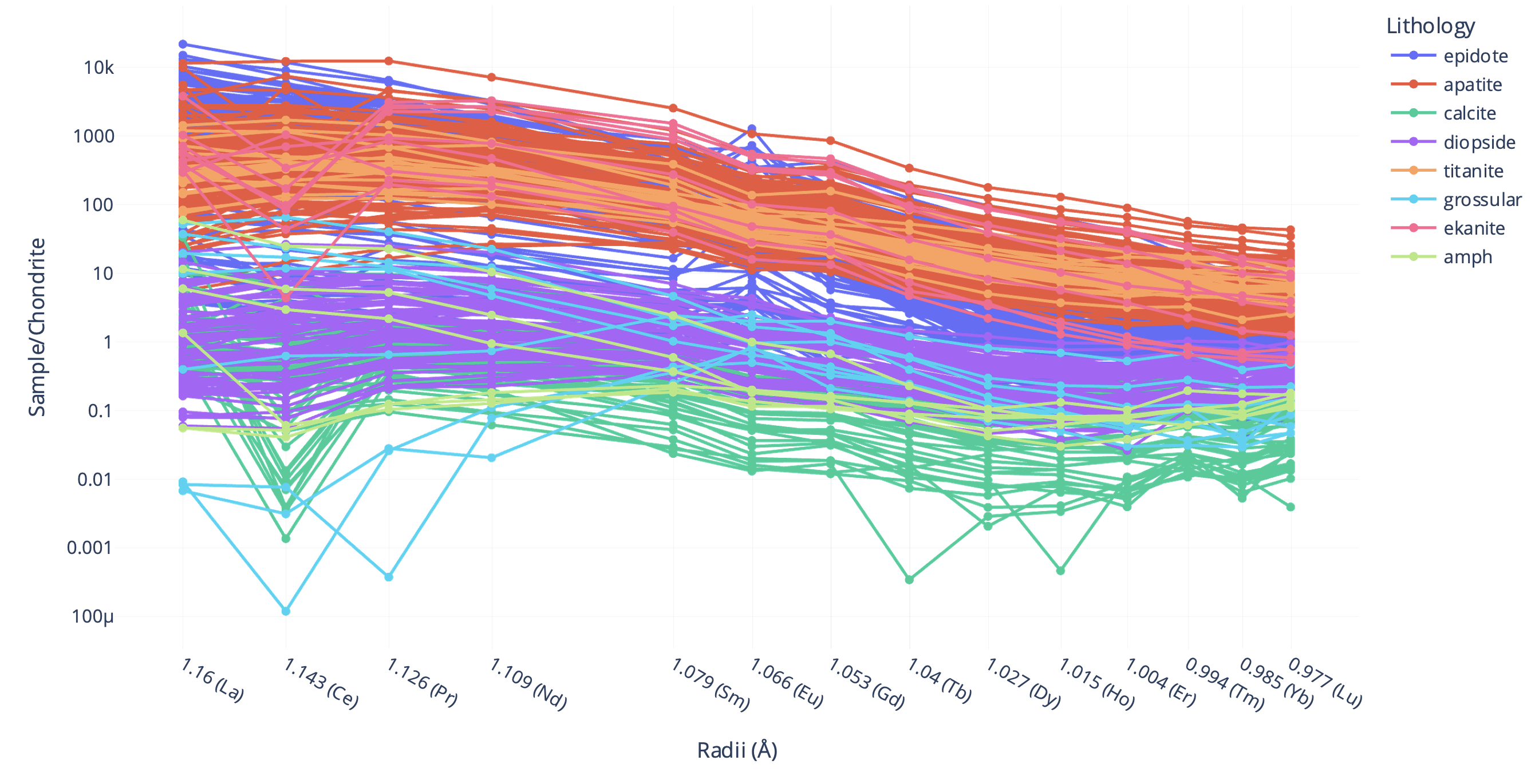}
\caption{Figure 7: Example spider diagram taken from the REEkit application. Colours in the diagram are associated with the mineralogy of each sample.}
\end{figure}

\paragraph{Scatter plot} Scatter plots (Figure 8) provide the fundamental framework for visualising data distributions, allowing for the representation of two variables and their relationships in a two-dimensional space. This versatile technique finds use in various fields and disciplines for the exploration and analysis of data patterns. In the context of this study, it will serve as an additional control method for the lambda data points.

\begin{figure}
\includegraphics[width=10cm]{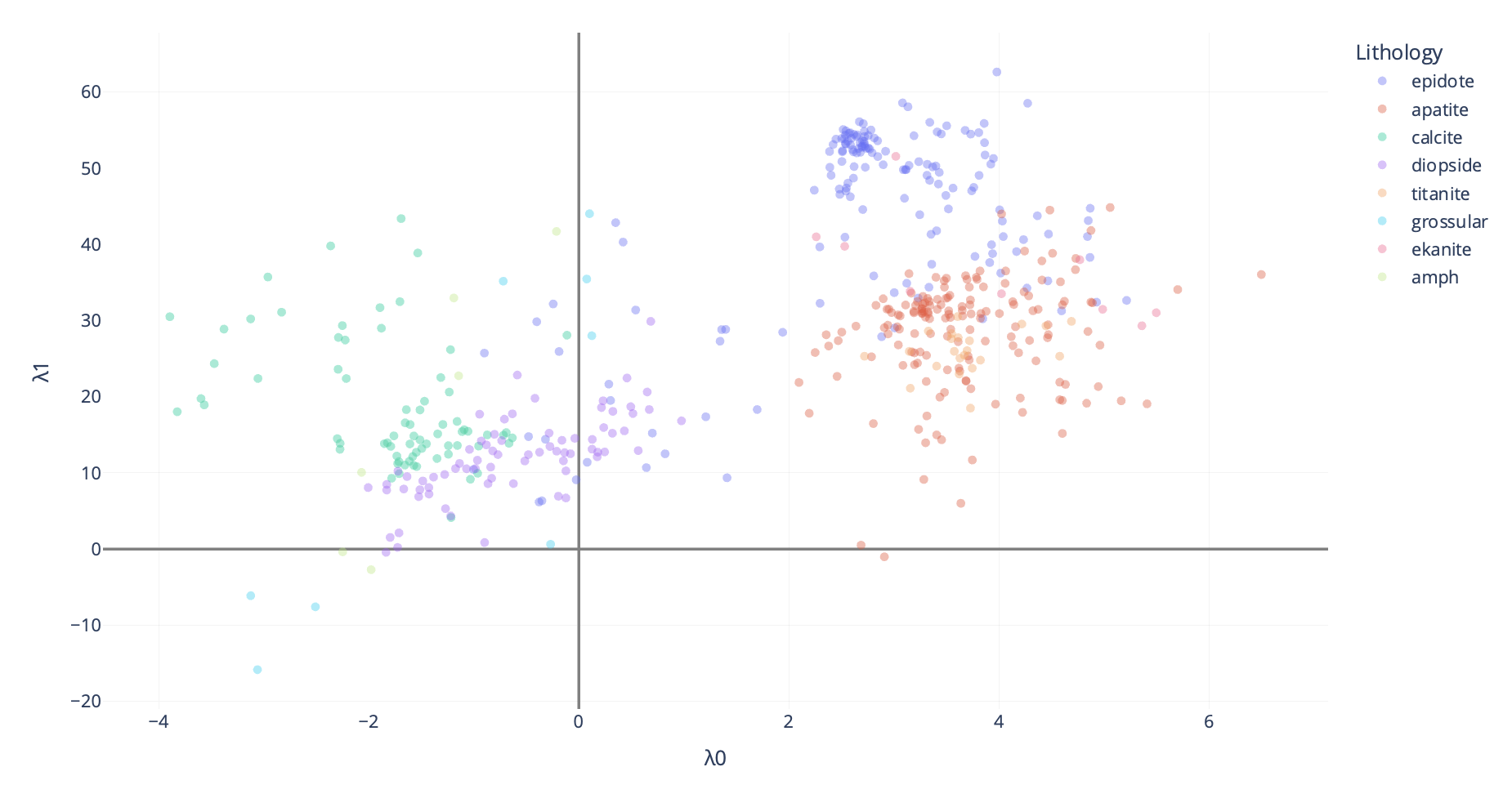}
\caption{Figure 8: Example scatter plot taken from the REEkit application. Colours in the diagram are associated with the mineralogy of each sample.}
\end{figure}

\paragraph{3D scatter plot} 3D scatter plots (Figure 9) are beneficial for visualising relationships among three numerical variables in a single plot, especially for exploring complex interactions and correlations. They provide depth and perspective, revealing hidden patterns within data, which is especially useful for multidimensional datasets (Busstra, Hartog and Van ’T Veer, 2005). Additionally, they can help identify clusters or groups of data points that may not be easily visible in traditional 2D scatter plots.

Traditionally, 3D visualisation in mineral exploration has primarily been applied within a geospatial context (Jackson, 2010). However, visualising data in multiple dimensions can reveal intricate patterns and trends which are overlooked in 2D. This capability holds significant relevance for lambda data, justifying its incorporation into this study.

\begin{figure}
\includegraphics[width=10cm]{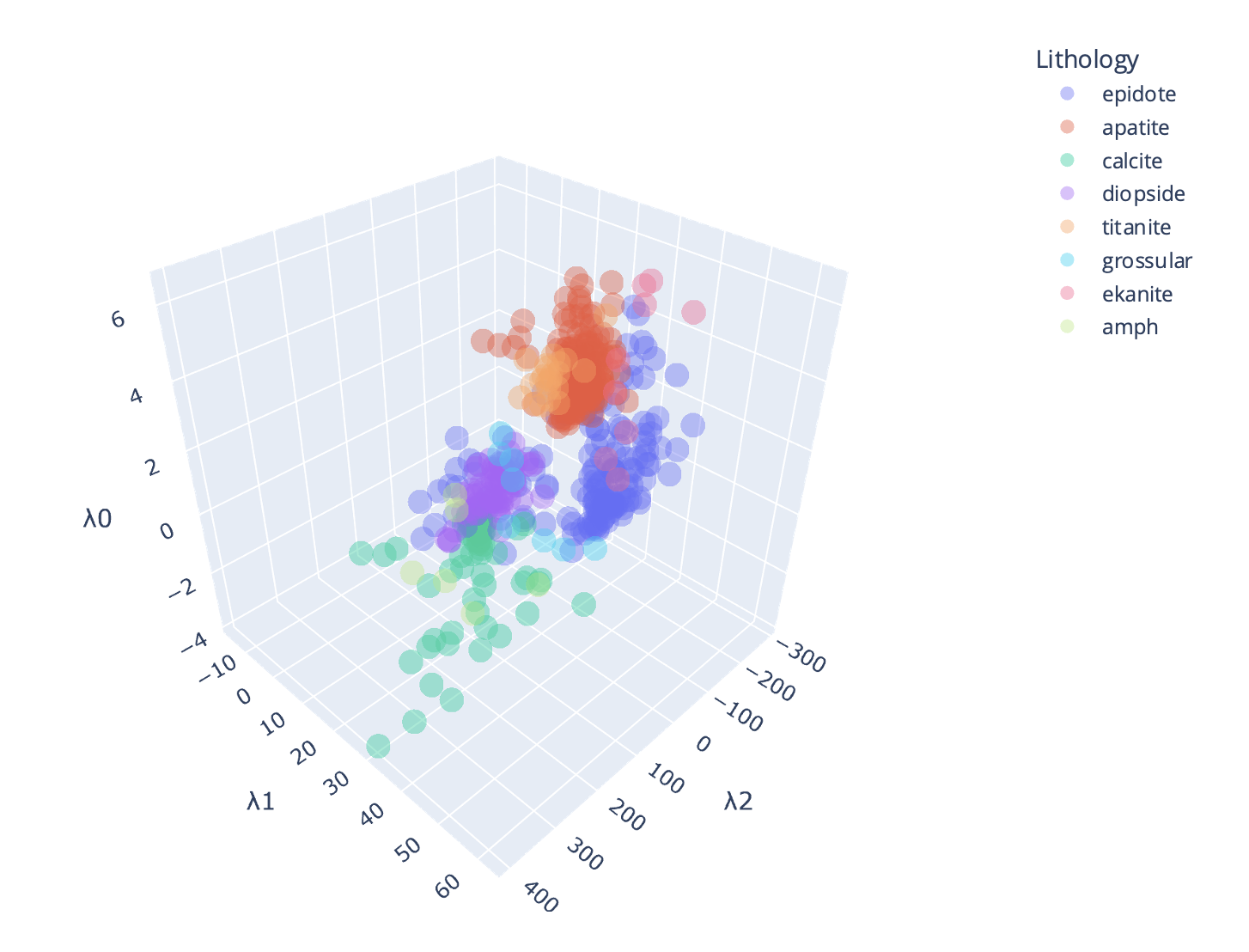}
\caption{Figure 9: Example 3D scatter plot taken from the REEkit application. Colours in the diagram are associated with the mineralogy of each sample.}
\end{figure}

\paragraph{Scatter plot matrix} The scatter plot matrix visualisation technique (Figure 10) facilitates the pairwise comparison of distinct variables, and for this reason, it finds extensive utility in the field of geochemistry for assessing correlations among various geochemical parameters (Grünfeld, 2005). 

This visualisation method is significant and worth inclusion in this study because it enables the exploration or relationships between multiple variables in a 2D context, which often constitute the most intriguing aspects of geochemical analysis (Kononenko and Kukar, 2007).

\begin{figure}
\includegraphics[width=10cm]{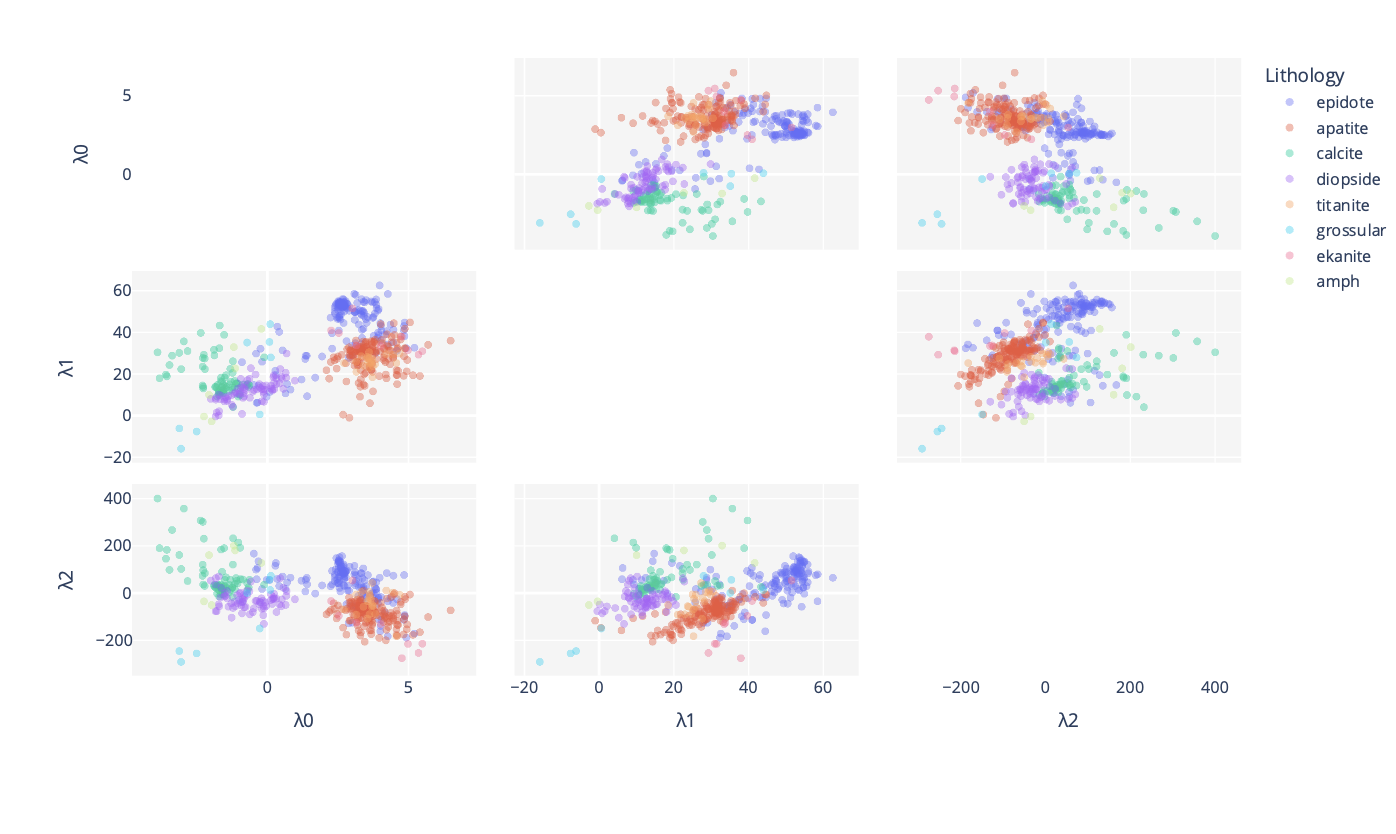}
\caption{Figure 10: Example scatter plot matrix taken from the REEkit application. Colours in the diagram are associated with the mineralogy of each sample.}
\end{figure}

\paragraph{Density contour plot} The use of density contour plots (Figure 11) to represent data density has proven to be a valuable technique for exploring extensive datasets. This holds true in a general context and is particularly prevalent within the field of geochemistry, primarily within geospatial and mineral resource estimation contexts (Agnerian and Roscoe, 2002). Density contour plots work by visually representing the distribution of data by showing the frequency of different values occurring within specific intervals (contours). These visualisations help in understanding the underlying patterns and variations in the data, providing insights into a datasets central tendency, and spread.

When the base density contour visualisation was combined with additional marginal visualisations, such as histograms (Figure 11a) and rug plots (Figure 11b), there was additional opportunity to gain valuable insight for the data. The accessory plots visually display data distribution on the x and y axes, utilising histograms to group data into intervals and rug plots to represent individual data points with dashes. It's worth noting that many of the tools discussed in section 2.3 offer the capability to create these plots, making it a fitting choice to include the density contour method into this study.

\begin{figure}
\includegraphics[width=10cm]{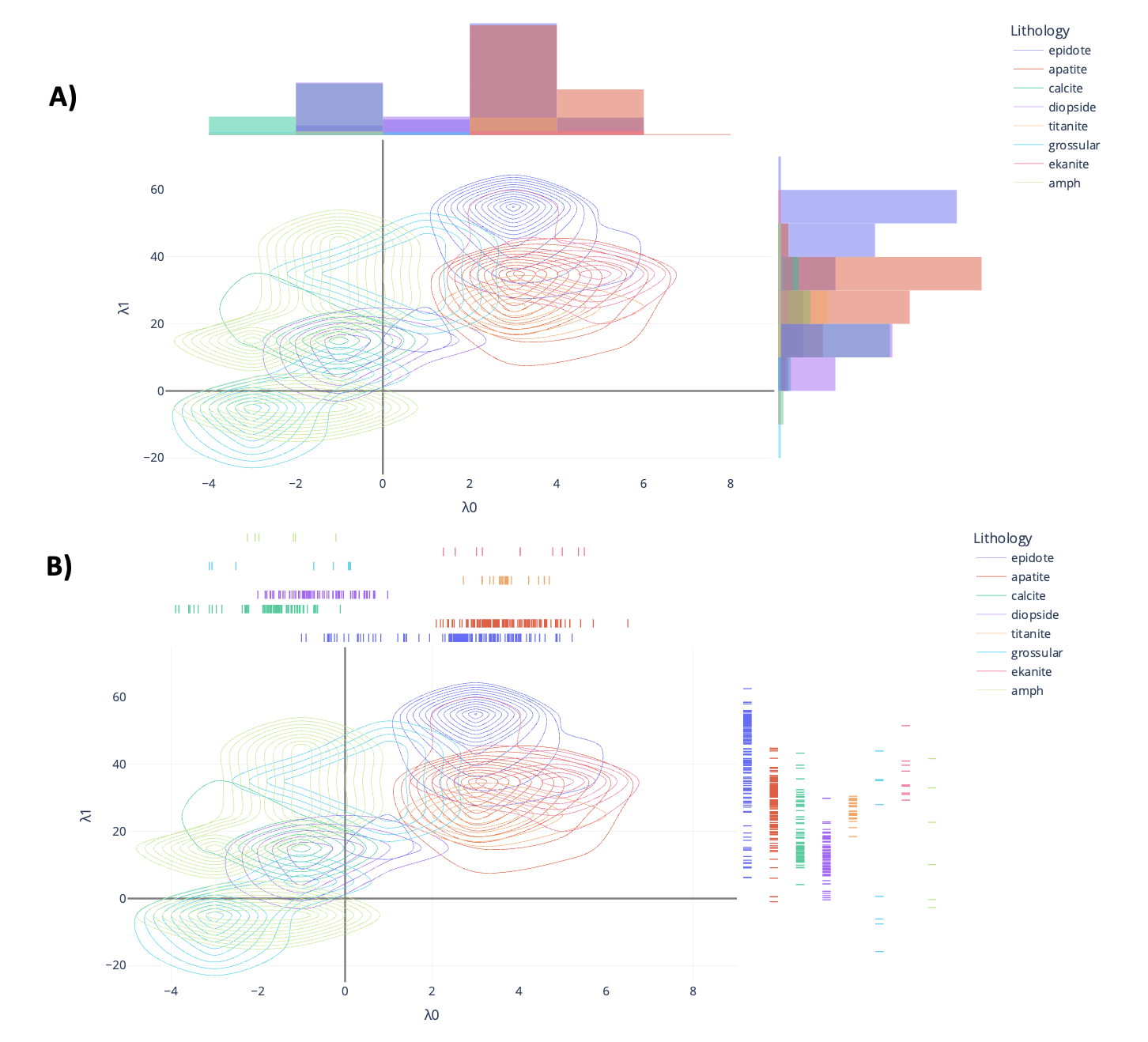}
\caption{Figure 11: Example density contour plots taken from the REEkit application. Colours in the diagram are associated with the mineralogy of each sample. (A) Showing the \textbf{histogram} accessory visualisations along the x and y axis; (B) Showing the \textbf{rug} accessory visualisations along the x and y axis}
\end{figure}

\paragraph{Violin plot} Violin plots (Figure 12) are valuable tools in geochemical analysis as they facilitate the comparison of distribution of data across multiple variables or categories. By incorporating the density tracing alongside a box plot, violin plots offer a more precise representation of the distribution's shape. This enhanced depiction includes the ability to reveal distinct data clusters, accentuating peaks, valleys, and bumps within the distribution (Hintze and Nelson, 1998). 

This visualisation method has unique benefits for identifying trends, outliers, or clusters within geochemical data, making it an important visualisation method to assess as part of this study.

\begin{figure}
\includegraphics[width=14cm]{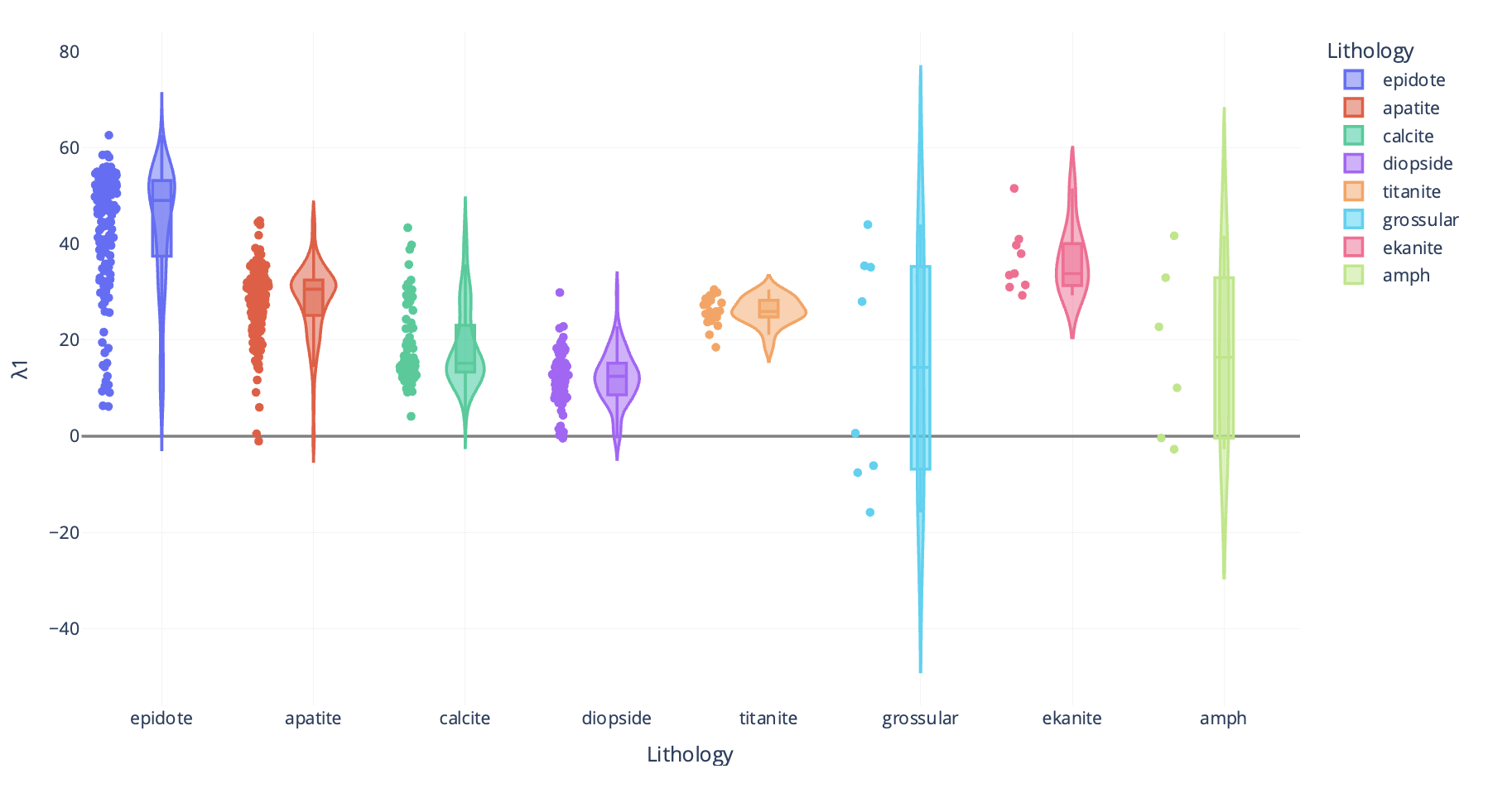}
\caption{Figure 12: Example violin plot taken from the REEkit application. Colours in the diagram are associated with the mineralogy of each sample. Each violin corresponds to a specific mineral category, displaying its distribution through a box plot for statistics, a surrounding trace indicating density, and individual points representing actual values within the category.}
\end{figure}

\subsubsection{REEkit design and development}

REEkit is an advanced single page web application designed to process REE geochemical data in CSV format. Upon receiving the input data, the application performs calculations to determine lambdas for each specified pattern. These calculated values are then visualised using the set of visualisations described in section 3.1.1.

The evolution of the REEkit web application underwent multiple stages. In its initial iteration, developed using R and the R Shiny framework, it featured a simple front-end dashboard interface. While R offered a wide array of visualisation libraries and a robust statistical engine, the Shiny framework had limitations in providing user-friendly interface components and delivering customisable and interactive visualisations. Consequently, we shifted from R to Python for the second iteration of the app.

The second app iteration was built in Python using the Python Dash framework. Unlike R Shiny, Python provided better interactive visualisation support and a much broader selection of libraries with user interface widgets and elements. All changes to the application have been remotely tracked using Git version control and can be accessed publicly on GitHub (link available upon request to the author).

For the purpose of this study, REEkit was hosted on Google Cloud App Engine. This hosting solution allows participants to access the visualisation any time and from anywhere, making participant engagements during the study easier to conduct over Zoom.

\subsubsection{Functionality}

Discussion with stakeholders during the development phase of REEkit (the main interface is shown in figure 13) led to the incorporation of numerous additional features within the web application. These enhancements, which are outlined below, greatly increase the functionality and user experience of the platform.

\begin{figure}
\includegraphics[width=14cm]{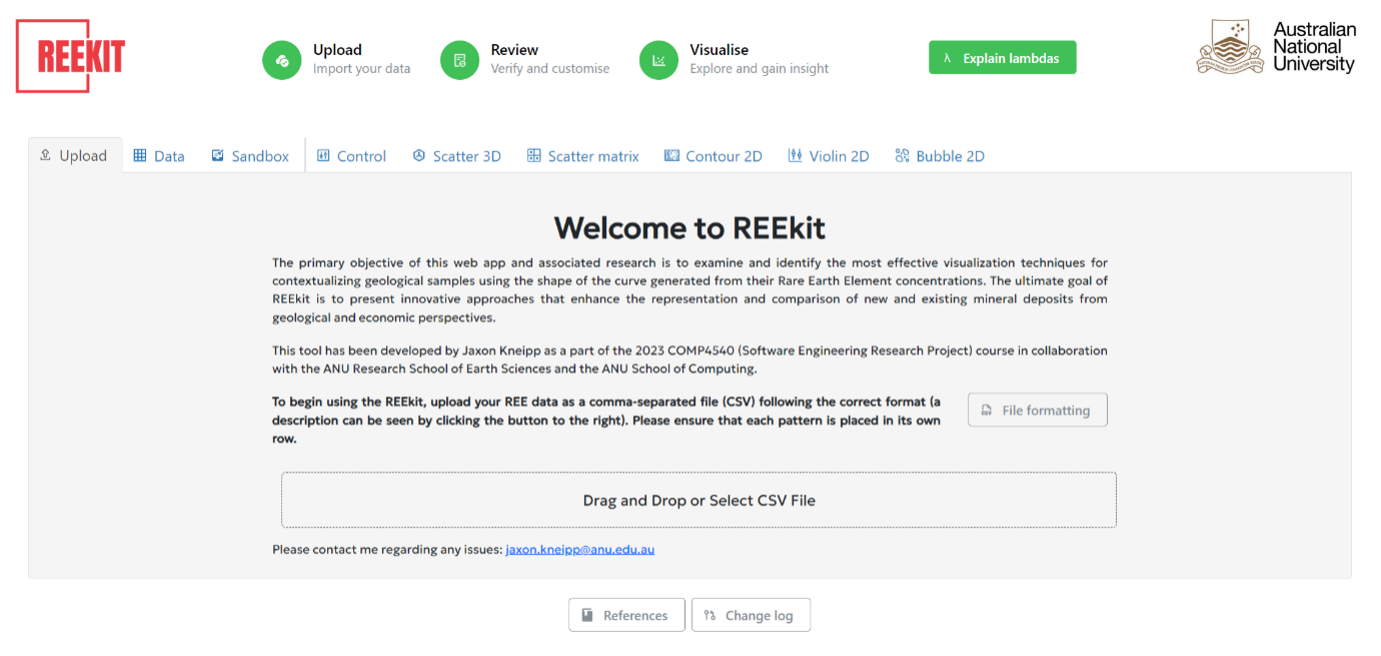}
\caption{Figure 13: Screenshot of the welcome interface for the REEkit application.}
\end{figure}

\paragraph{Tutorial} When users initially upload data to the website, they encounter a user-friendly tutorial that explains the concept of lambdas and their practical applications in data analysis. This instructional feature was incorporated based on insights gathered during the initial requirements gathering phase of the application's development, in which numerous potential users revealed their limited understanding of lambdas and their utility.

\paragraph{Sandbox tool} In addition to featuring a tutorial screen designed to introduce and simplify the concept of lambdas, a sandbox tool was developed (Figure 14). This tool allows users to visually explore the impact of different patterns on lambda values and vice versa. It offers full customisation options and plays a crucial role in highlighting and demonstrating the transformation between patterns and the data imported into the application. User can also view any of their own patterns in the sandbox by clicking on a point in certain visualisations then clicking ‘Open in Sandbox’.

\begin{figure}
\includegraphics[width=14cm]{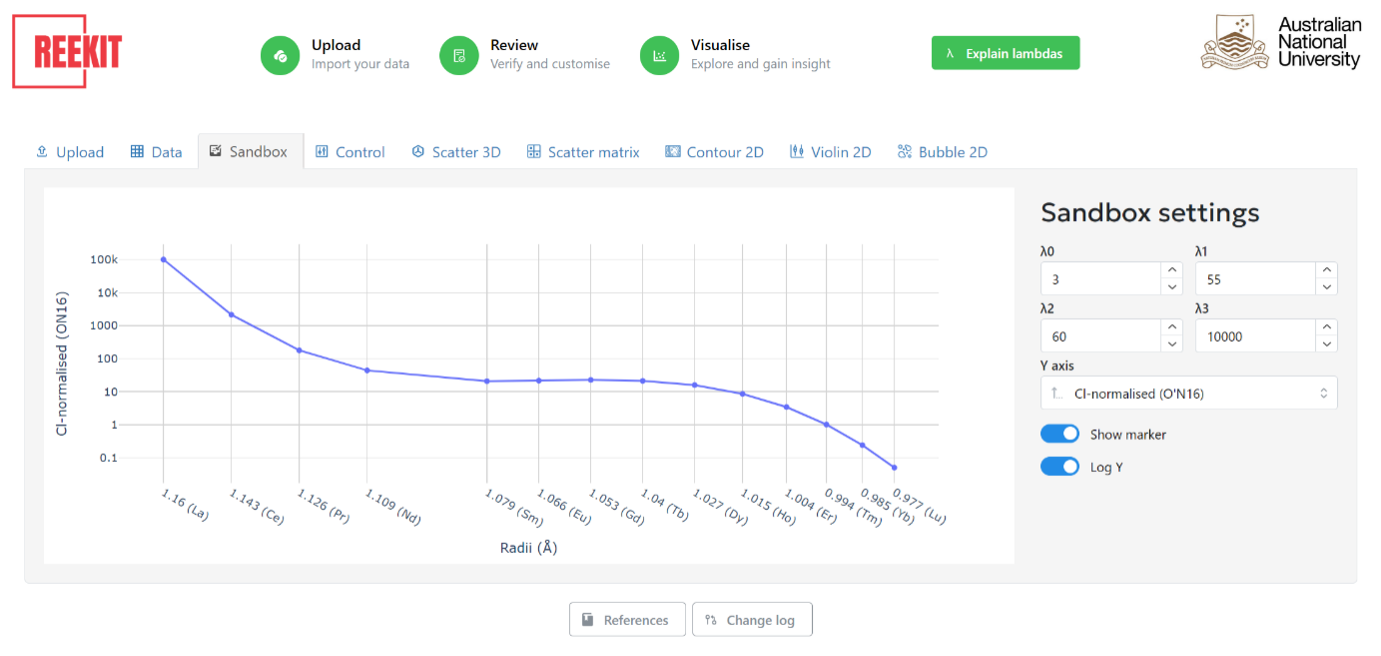}
\caption{Figure 14: Screenshot showing the Sandbox tool from REEkit.}
\end{figure}

\paragraph{Visualisation customisation} When users hover over a figure generated with plotly.py, a convenient toolbar (referred to as the 'mode bar' within the framework) appears in the top-right corner of the figure. This feature provides users with a range of options for seamless interaction with the visual content (‘Configuration in Python’, 2023). This toolbar allows users to engage with the visualisations by enabling actions such as panning, zooming, and region selection. Additionally, users have the ability to download PNG copies of each visualisation, enhancing the utility of the generated content.

\subsection{Participants}

\subsubsection{Participant demographics}

The study's participant demographic includes potential users, stakeholders, and domain experts, consisting of 10 individuals actively engaged in the global mineral exploration industry, reflecting the study's broad scope. Their backgrounds span academia, industry, and government sectors, ensuring a diverse range of perspectives. Participants also have regular interaction with REE data enhances their relevance to the study.

All participants possessed tertiary education, providing a strong foundation for their contributions to the study. Their collective experience encompasses a diverse range of commodities, including lead-zinc, gold, uranium, and of course REEs, demonstrating a detailed understanding of the workflow and processes involved in global mineral exploration. 

Importantly, the participant group encompassed various cultural backgrounds and genders. This aspect adds depth to the study's insights and contributes to the overall strength of its outcomes.

\subsubsection{Sampling strategy}

Participants were identified and selected through a combined approach, incorporating both convenience sampling and the snowball sampling method (Elfil and Negida, 2017). Convenience sampling involved reaching out to easily contactable individuals within the mineral exploration industry who met the study's criteria. Subsequently, the snowball sampling method was employed, wherein some initial participants assisted in identifying and referring other qualified individuals, thus broadening the pool of participants.

\subsubsection{Participant selection criteria}

Individuals participating in the study were required to possess experience within the mineral exploration sector, with a specific emphasis on the field of REEs. Additionally, participants were expected to have regular involvement with REE data. 

It's important to note that individuals with substantial visual impairment were not included in the study. This decision was driven by the study's primary focus on evaluating visualisations, which necessitated participants' ability to engage with visual content effectively.

\subsubsection{Sample size rationale}

The goal of a qualitative approach to visualisation evaluation research is to gain in-depth insights and understanding from users rather than achieving statistical generalisability through experimentation. 

As such, sample size is not determined by traditional statistical calculation but by the concept of data saturation or, the point where no new information or insights are emerging from additional participants or data collection (Saunders et al., 2018). Literature documenting user experience through contextual inquiry have reported achieving data saturation with sample sizes ranging from 6---12 participants (Ladner, 2016).

Considering the exploratory nature of this study and the intention to uncover detailed insights through means of contextual inquiry, a sample size of 10 participants is deemed appropriate. This number allows for a manageable workload in terms of data collection, transcription, and analysis, while still ensuring that a variety of perspectives and experiences are captured.

\subsubsection{Participant bias}

Participant bias refers to the tendency of participants to alter their behaviour, attitudes, or responses due to their awareness of being observed. To minimise levels of bias in this study, several strategies were implemented during the interview process:

\begin{enumerate}
\item \textbf{Clear explanation of study purpose}: the study is explained to participants prior to observing them. Conducting the engagement in a friendly and genuine manner helped to build repour and disarm participants. By establishing a comfortable environment, participants were more likely to engage candidly during the interview.
\item \textbf{Emphasised confidentiality}: Confidentiality was communicated to participants and held throughout all stages of the study. See section 3.5 for more details.
\item \textbf{On-going member checking}: this was done throughout the interview to ensure that the researcher and participant were both on the same page when it came to thoughts and points discussed during the interaction.
\end{enumerate}

By combining these strategies, the study sought to create an environment that encouraged participants to act naturally, express genuine attitudes, and provide authentic responses. These measures were instrumental in enhancing the validity and reliability of the research findings by reducing the impact of participant bias.

\subsection{Data collection}

Qualitative data was collected by contextual inquiry through participant interview and observation. Participant engagement ranged in duration from 45-90 minutes and were conducted in two parts – a semi-structured interview and an observational session where participants were supervised interacting with the REEkit platform and associated visualisations (outlined in section 3.1.1) used a series of three different datasets:

\begin{enumerate}
\item Geochemistry data of Nolans Bore obtained by LA-ICP-MS (Anenburg, 2019)
\item Geochemistry data of Mid-Oceanic Ridge Basalts (O’Neill, 2016)
\item Geochemistry exploration results (prepared data taken from public ASX announcement M24:ASX)
\end{enumerate}

The datasets were chosen because they provide various instances of REE geochemical data that can be utilised for creating visualisations. The key distinguishing factors among the datasets were the inclusion of mineralogical data (dataset 1), information about the geological formation environment (dataset 2), and more detailed sample description for each pattern (dataset 3). It is important to note that biases including participant preference for certain categorisations (such as or hole ID or lithology) may have affected their responses and attitudes towards the suite of visualisations. 

During the interview portion of the engagement, participants were asked questions pertaining to their experience in the exploration industry as well as their experience using visualisations and interpreting REE data in their line of work. The information collected during this portion of the study was useful during the data analysis stage when correlating current work patterns with visualisation effectiveness. 

Following the interview, participants were asked to use the provided visualisations to accomplish several simple tasks using a talk-aloud protocol. The tasks included:

\begin{itemize}
\item Identifying any trends or correlation in the provided dataset
\item Identifying any anomalies in the provided dataset
\item Identifying any geochemical errors in the data being presented
\end{itemize}

Using these tasks as a guideline, participants would verbalise their thoughts, feelings, and reactions towards the different visualisations as they interacted with the REEkit tool. Throughout the engagement, handwritten notes on comments or observations the participant made were taken. 

On completion of the interview, the notes were collated, digitised, and expanded upon using the recordings from the session, for later analysis (discussed in more detail in section 3.4). 

\subsubsection{Procedure}

The data collection process in this study adhered to the following procedure:

\begin{enumerate}
\item \textbf{Planning}: A versatile discussion guide was designed to align with research goals, incorporating open-ended questions and ensuring cultural sensitivity. Furthermore, approval from the ANU Human Research Ethics Committee was obtained to uphold ethical research standards, encompassing informed consent, privacy protection, participant well-being, and ethical compliance.
\item \textbf{Recruitment and consent}: Participants were recruited using the methods outlined in section 3.2.2. Participants were provided with an explanation of the study's purpose, procedures, and potential benefits and risks (see Appendix A for a copy of the Participant Information Sheet). Written informed consent was then obtained (see Appendix B for a copy of the consent form).
\item \textbf{Contextual interviews}: Engagements were conducted via Zoom and recorded for subsequent analysis and transcription. Participants were encouraged to share their thoughts, feelings, and decision-making processes with interacting with the different visualisation in the REEkit application.
\item \textbf{Questioning}: During the engagement, open-ended questions were asked to probe deeper into participants' experiences. For example:
\begin{itemize}
\item "In what way could this visualisation technique integrate into your workflow?"
\item "How do these visualisations help convey information not readily available?”
\item "What aspects of the data presentation in this visualisation do you find useful or not?
\end{itemize}
\item \textbf{Data collection}: Notes taken during the engagement were initially recorded on a physical discussion guide (see Appendix C for a simplified version of the discussion guide) during the session. Subsequently, they were digitised and enriched after each interview concluded.
\end{enumerate}

\subsection{Data analysis}

The data analysis procedure encompassed the extraction of relevant data points and comments from the data collected. These data points were subsequently categorised into coherent thematic clusters, facilitating the identification of underlying patterns and relationships between data collected.

These thematic clusters formed the basis for the findings presented in this report. These insights were related to the research objectives and presented as the primary output of this research project. This approach aligns with the analytical approach presented by Kerren et al. in their textbook ‘Evaluating Information Visualizations’ (Kerren et al., 2008). The findings from this study are discussed in more detail in section 5.

\subsection{Ethical considerations}

In this study, ethical considerations were essential in order to safeguard the well-being and confidentiality of participants and ensure their protection and privacy throughout the research process. Firstly, participation was entirely voluntary, and individuals had the right to decline involvement or withdraw from the study at any point, without needing to provide an explanation. 

Secondly, strict confidentiality measures were put in place to protect the identities of participants, with the option for anonymity. Compliance with the Australian Privacy Act of 1988 further underscores the projects commitment to safeguarding personal information. 
Overall, the research emphasised non-interference, making it clear that participation was optional and would not impact participants' professional affiliations or relationships. These ethical considerations prioritise participant autonomy, confidentiality, informed decision-making, and voluntary consent, while ensuring a responsible and ethical research environment.

\section{Results}

The study's findings are centred on the practical applications of the visualisation methods (outlined in section 3.1.1) in mineral exploration. The visualisations demonstrated their usefulness in communicating with stakeholders and revealing the spatial distribution of different data groups based on participant responses. They also aided in understanding how these groups change when categories like or depth are modified.

In this section, a more comprehensive examination of each of these findings is provided. The aim is to address the overarching research objectives of identifying specific contexts where lambda data offers enhanced insights. Determining optimal visualisation methods for isolating lambda data is a key focus and exploring integration techniques for deeper geological insights utilising multiple visualisation methods.

\subsection{Software Tooling: Current practices and the adoption of Lambdas}

Participants mentioned using a diverse range of software tooling when analysing and visualising geochemical REE data in their current line of work. The most commonly mentioned tool was Excel and other spreadsheet software such as Numbers and Calc. Several more specialised software tools were also mentioned including ioGAS (Figure 15), QGIS analytics, Micromine and Leapfrog, with only a small subset of participants highlighting A/BLambdaR as a tool they had used to visualise their data. 

Perceptions about these specialised tools were varied, with some users appreciating their role in streamlining the data analysis process, while others criticised them for imposing limitations and constraints on their individual workflows. Additionally, several participants pointed out licensing issues and the high costs associated with the current tools.

\begin{figure}
\includegraphics[width=14cm]{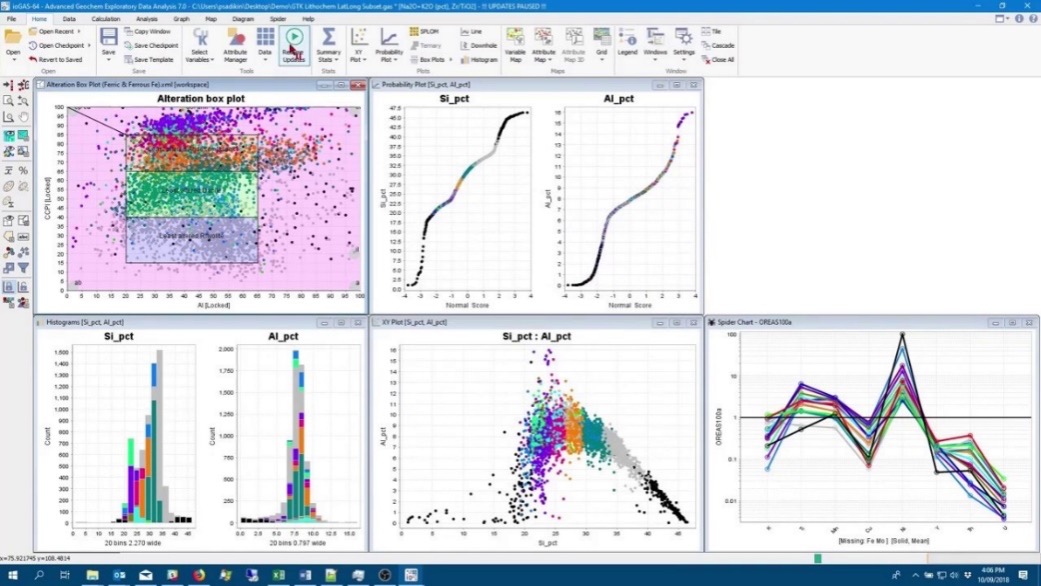}
\caption{Figure 15: Screenshot showing the ioGAS interface (Lawie and Sadikin, 2018)}
\end{figure}

Participants, especially those specialised in the field of REE exploration, including areas like heavy rare earth elements (HREE), mineral sands, or carbonatite style deposits, expressed challenges in finding suitable visualisation tools tailored to their needs. Many of them resorted to using scientific programming languages like Python and R to create custom visualisations that catered specifically to their unique requirements, a process which takes considerable time and effort.

The absence of a standardised workflow for geochemical data analysis among participants became evident through diverse viewpoints regarding the applicability and usefulness of various visualisations within the participant pool, underlining the lack of consistency in their approaches.

In regard to the adoption of lambdas, most participants had heard of or toyed with the idea of parameterising REE patterns but not explored it in depth. Most academics working in the field of REEs had used lambdas in their research, but the technique had not been widely used by participants from industry and government due to the novelty of the lambda approach.

\subsection{Effectiveness of lambda visualisations}

All evaluated visualisations enhanced participants' ability to analyse the provided data by enhancing their capacity to identify patterns and distinct data populations. Notably, however, this was particularly apparent when utilising the lambda data visualisations compared to the control visualisation. Participants could clearly see that lambda data effectively simplified REE patterns into distinct points, facilitating easier separation and categorisation of data into potentially meaningful populations.

Interestingly, participants particularly found that visualisations such as scatter plots (2D and 3D) and density contour plots, made it easier to identify groupings of data in space. 

Participants typically used the visualisations to progress through the following steps as part of their data discovery: first, to identify distinct populations within the data; then, to analyse the patterns associated with these populations; and finally, to delve into the influence of underlying geochemical processes on these patterns and populations.

Participants noted that these workflows were common when dealing with geochemical data analysis for mineral exploration. Many participants highlighted that integrating lambda data with these visualisations could substantially expedite their analysis processes. 

Despite this, there was a widespread acknowledgment that the mathematics and derivation of the lambda values presented a significant cognitive challenge due to its complexity. Participants noted that comprehending the underlying significance of where data sat spatially required substantial mental effort and thoughtful consideration. In line with this, the 2D lambda visualisations were the preferred choice of many participants due to their simplicity. In contrast, the 3D scatter plot was often described as "overwhelming" and deemed difficult to interpret, primarily due to this complexity. 

\subsection{Visualisation method effectiveness and limitations}

\paragraph{Spider diagrams} (Figure 16) are a well-established and conventional visualisation method with many participants noting its familiarity and remarking that it offered no ground-breaking innovations. Consequently, its widespread adoption in industry prompted many to comment on its advantages for sharing and communicating findings to others.

Nevertheless, limitations were acknowledged, including issues related to information density and the inability to provide comprehensive insights into the intricate trends within the large datasets. Some participants found it valuable in cases where significant differences in patterns within the dataset were evident, but not when the differences were more subtle.

\begin{figure}
\includegraphics[width=14cm]{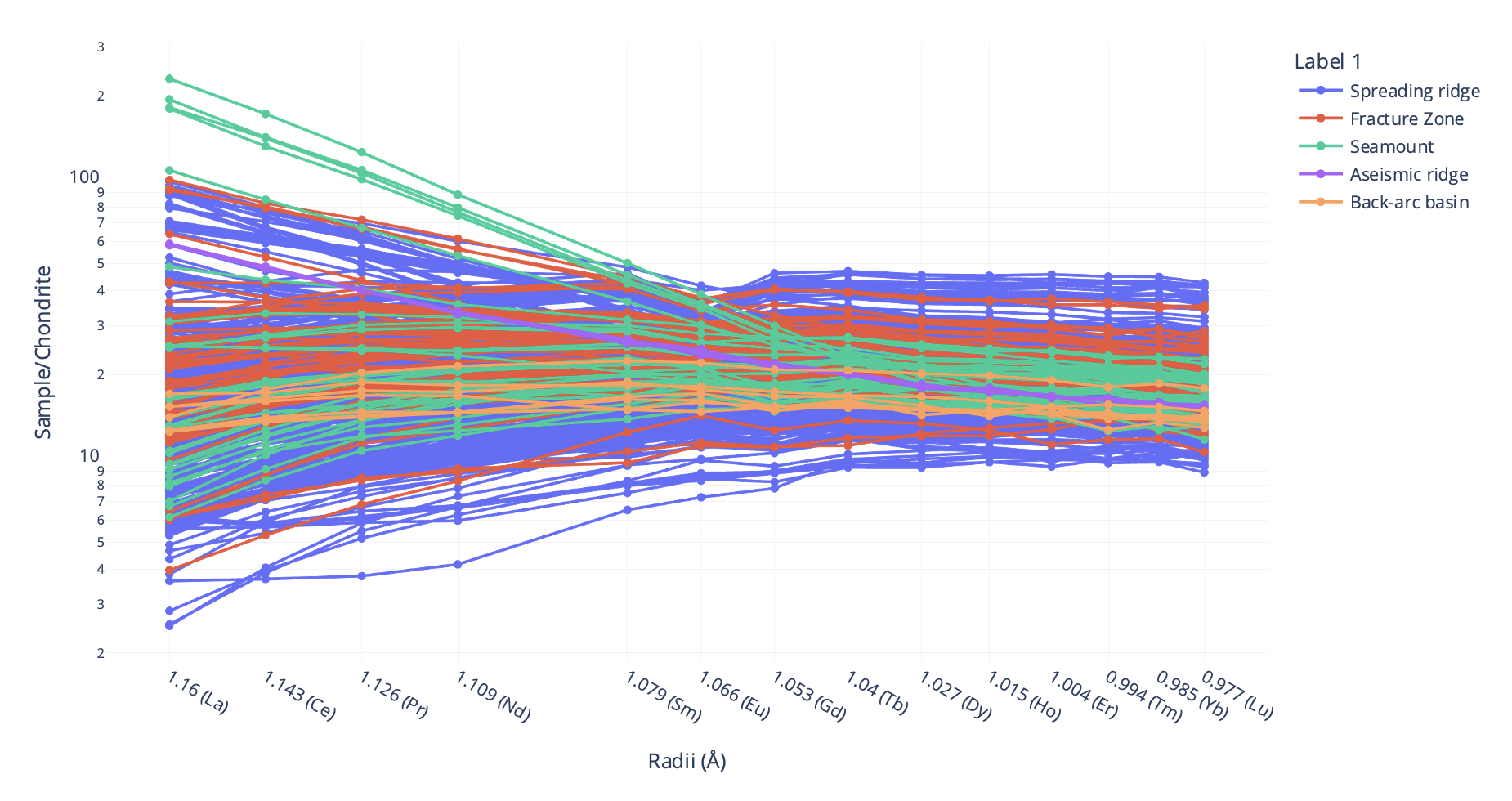}
\caption{Figure 16: Example spidergram taken from the REEkit application. The colours on the plot indicate the geological region of each sample.  Elements are represented on the x-axis by atomic radius in picometers, while the sample’s normalised concentration is plotted on the y-axis.}
\end{figure}

\paragraph{Scatter plot} (Figure 17) had several identified strengths, including the ability to show offsets between datasets effectively. The scatter plot was characterised as a simple visualisation, favoured by some participants. Participants saw it as an improvement over simply viewing the raw data but perhaps less effective at revealing dimensions and trends. Some respondents found it useful when paired with specific accessory data such as mineralogy. 
However, it received criticism for its restricted customisation options, particularly in terms of size and shape of data points. It was generally viewed as a good method to employ, but nothing ground-breaking.

\begin{figure}
\includegraphics[width=14cm]{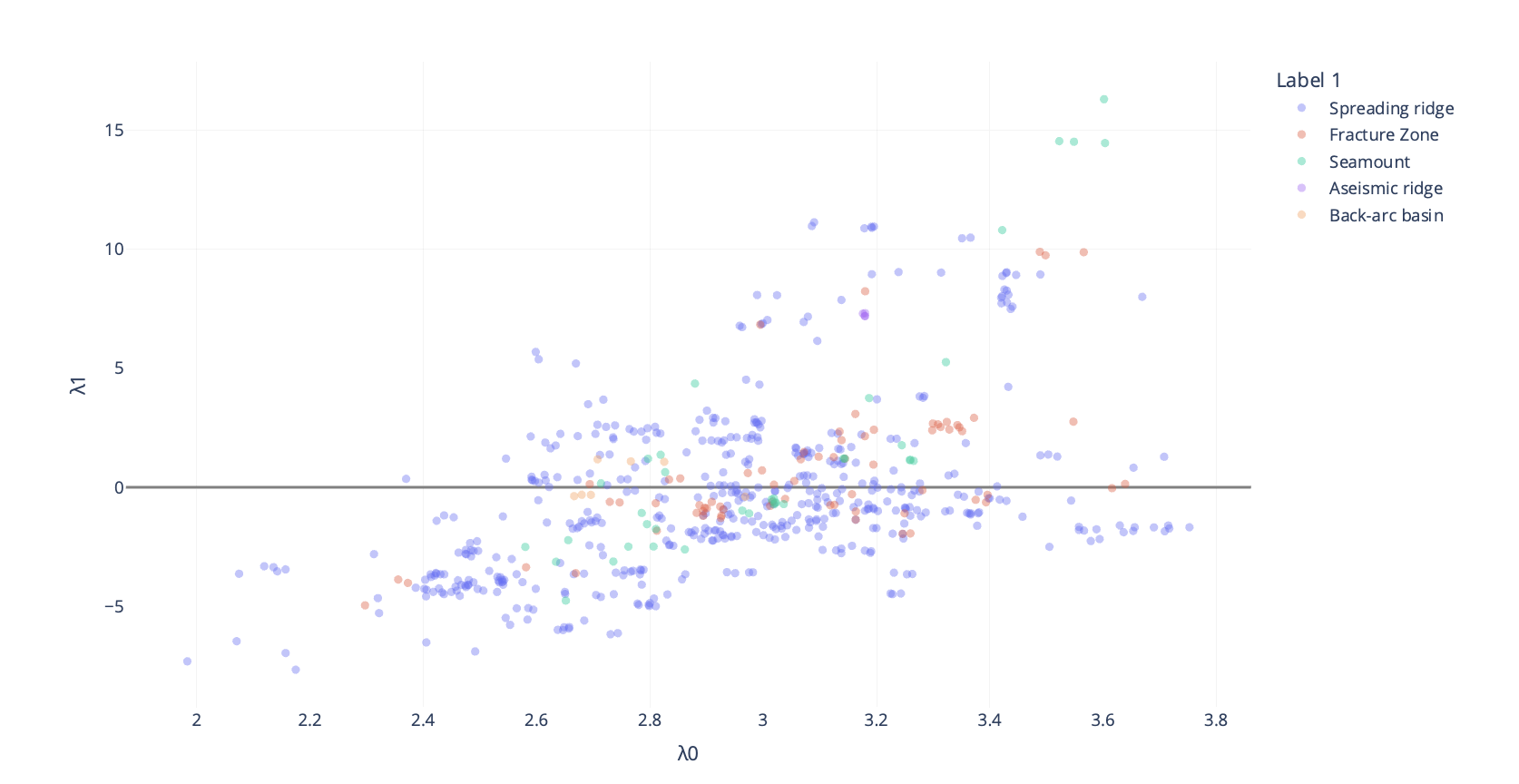}
\caption{Figure 17: Example scatter plot taken from the REEkit application. The colours on the plot indicate the geological region of each sample. The X-axis represents lambda 0 (REE abundance), and the Y-axis represents lambda 1 (heavy or light REE enrichment).}
\end{figure}

\paragraph{3D scatter plot} (Figure 18) allowed participants to quickly identify differences, anomalies, and trends within the provided datasets. Many noted its effectiveness in revealing discrete data point clusters (or populations) within large datasets, with colour coding playing a vital role in aiding pattern recognition. Nevertheless, some participants found this visualisation method challenging due to the complexity of interpreting multiple axes and the volume/distribution of data points in three-dimensional space.

\begin{figure}
\includegraphics[width=8cm]{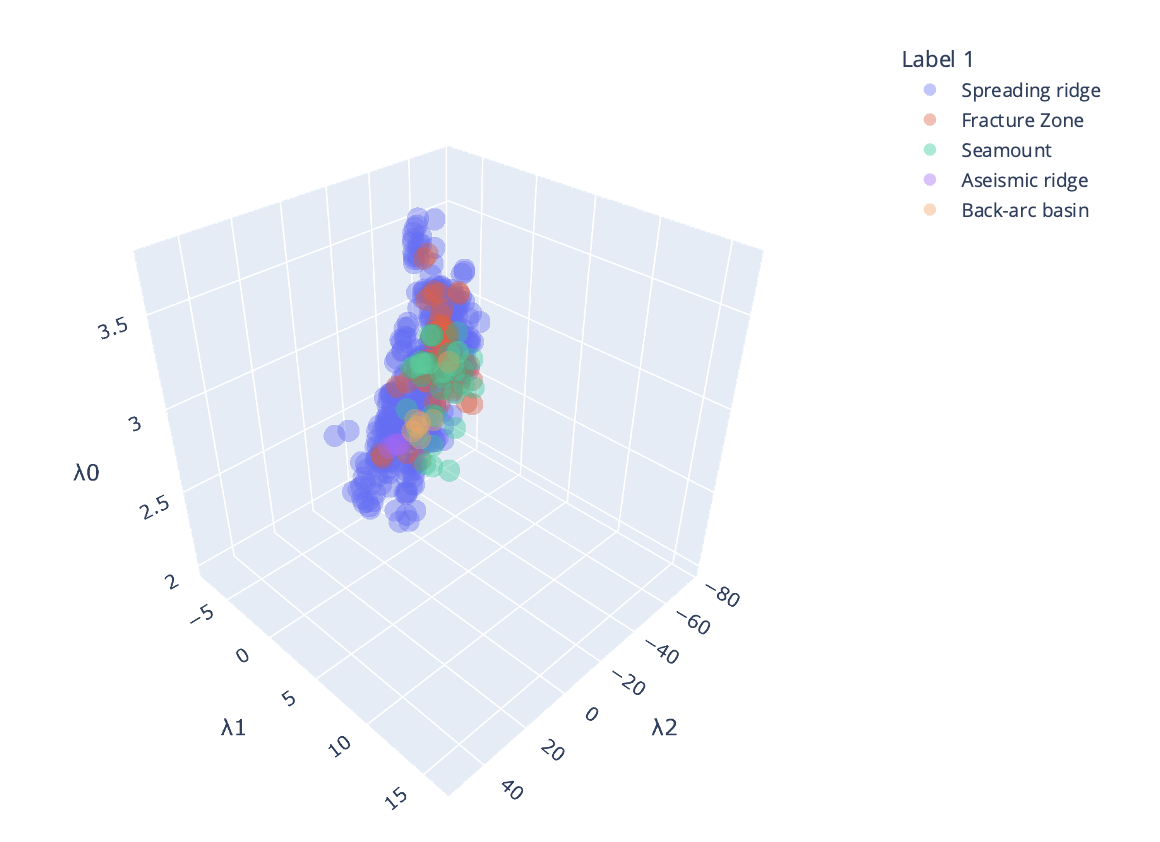}
\caption{Figure 18: Example 3D scatter plot taken from the REEkit application. The colours on the plot indicate the geological region of each sample. The X-axis represents lambda 2 (enrichment of middle REEs), the Y-axis represents lambda 1 (heavy or light REE enrichment) and the Z axis represents lambda 0 (REE abundance).}
\end{figure}

\paragraph{Scatter plot matrix} (Figure 19) excelled in providing a clear view of offsets among various data clusters, presenting information in an organised and comprehensive manner. Similarly juxtaposed with 3D Scatter plot by participants, it offers similar capabilities within a 2D space, making it more accessible for interpretation. Many appreciated the advantages of 2D visualisation, as it facilitated seamless sharing and communication of findings, whether through research papers or among team members during alternating swings at an exploration or mining site.

However, participants also noted complexities and a feeling of being overwhelmed, with many commenting on the high data density and the visualisations' intricate nature, which initially hindered their ability to interpret the data.

\begin{figure}
\includegraphics[width=8cm]{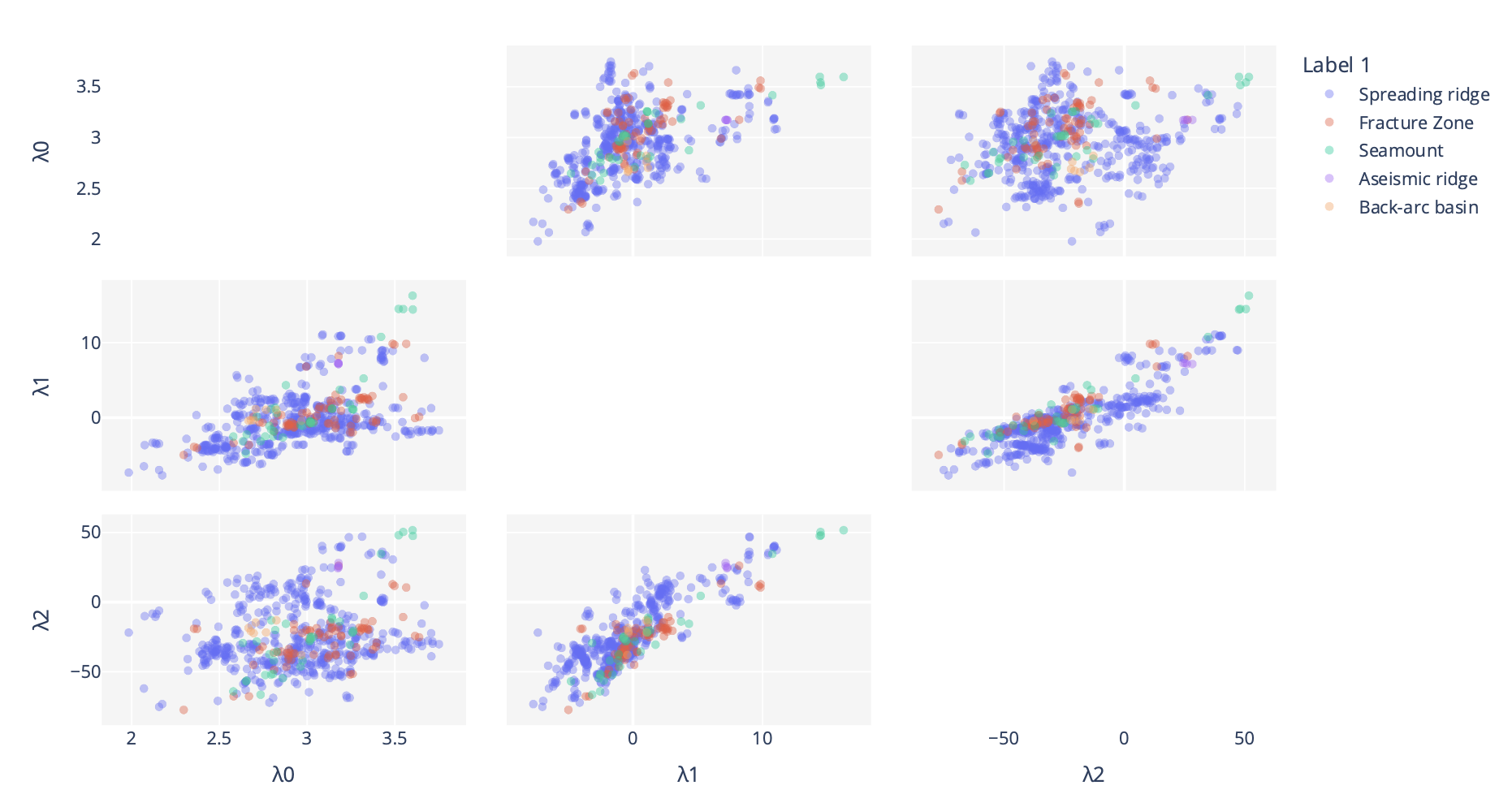}
\caption{Figure 19: Example scatter matrix plot taken from the REEkit application. The colours on the plot indicate the geological region of each sample. Lambda 0-2 are shown on both the X and Y axis.}
\end{figure}

\paragraph{Density contour plot} (Figure 20) was found to be a straightforward method for visualising trends and gaining a broader understanding regarding the distribution of data points across various categories within the dataset. Participants generally valued the simplicity of the concentric model, especially when handling larger datasets. 

However, several found it challenging to grasp the meaning behind the contours, considering it unintuitive. Moreover, some participants expressed dissatisfaction regarding the density of lines, mirroring concerns raised about information overload in spider diagrams.

Participants generally valued the additional distribution information displayed on each axis, with a general preference for the rug plot due to its ability to offer higher-resolution details on information simplified into density contours.

\begin{figure}
\includegraphics[width=8cm]{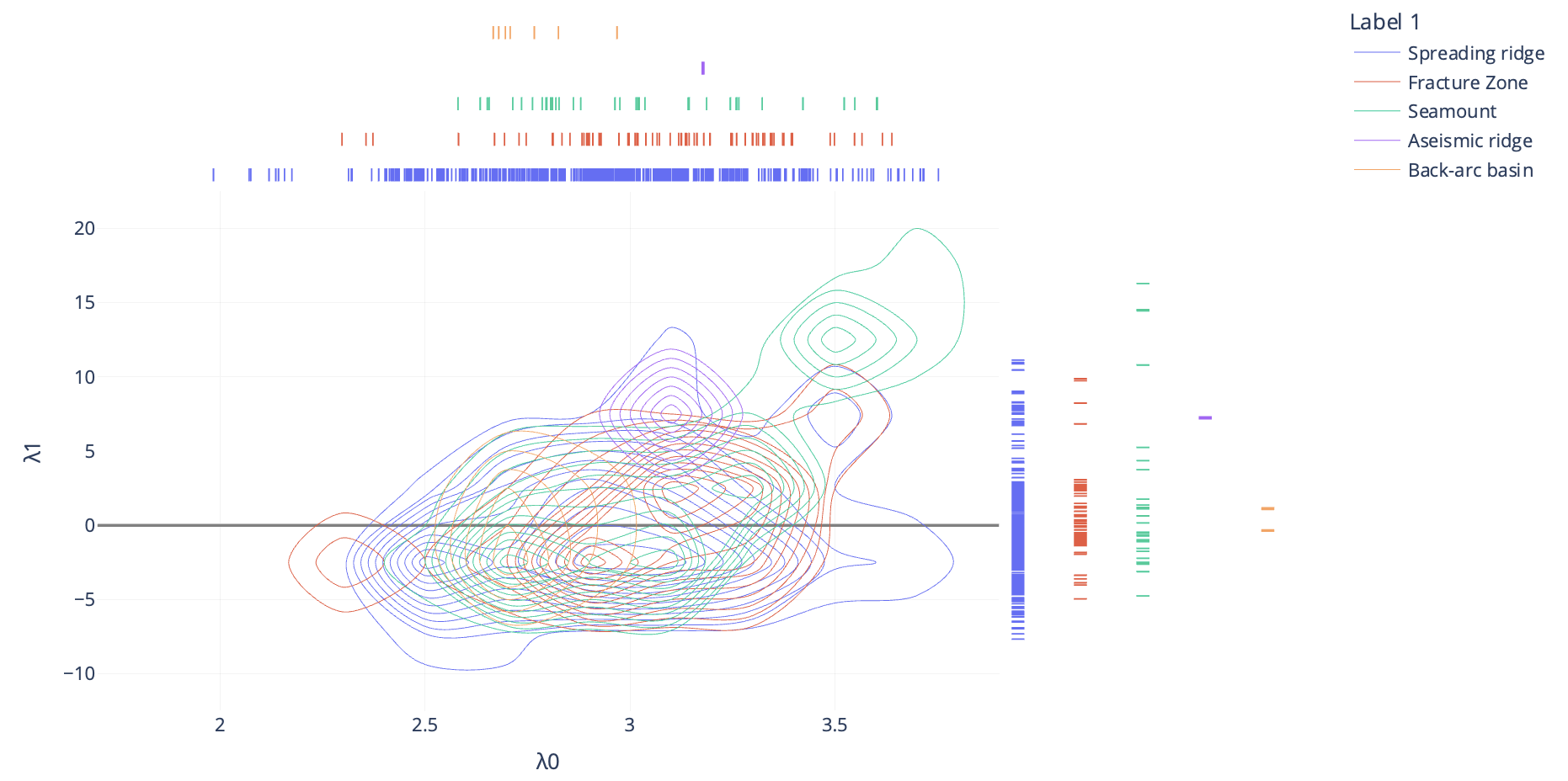}
\caption{Figure 20: Example density contour plot taken from the REEkit application. The colours on the plot indicate the geological region of each sample. The X-axis represents lambda 0 (REE abundance), and the Y-axis represents lambda 1 (heavy or light REE enrichment). The rug accessory plot is shown on both the X and Y axis.}
\end{figure}

\paragraph{Violin plot} (Figure 21) was powerful for univariate analysis within categorical data, such as or specific drill holes. Many participants noted that this visualisation method hadn't gained widespread adoption in the industry, primarily finding use in niche applications related to geochemistry. 

Typically, participants found this method to be most valuable as a supplementary visualisation when used in conjunction with scatter plots or density contours, especially for identifying distributions and trends within categories. 

Notably, several participants were able to pinpoint inaccuracies within a provided dataset using this visualisation, a detail that often went unnoticed when using other visualisation techniques.

\begin{figure}
\includegraphics[width=8cm]{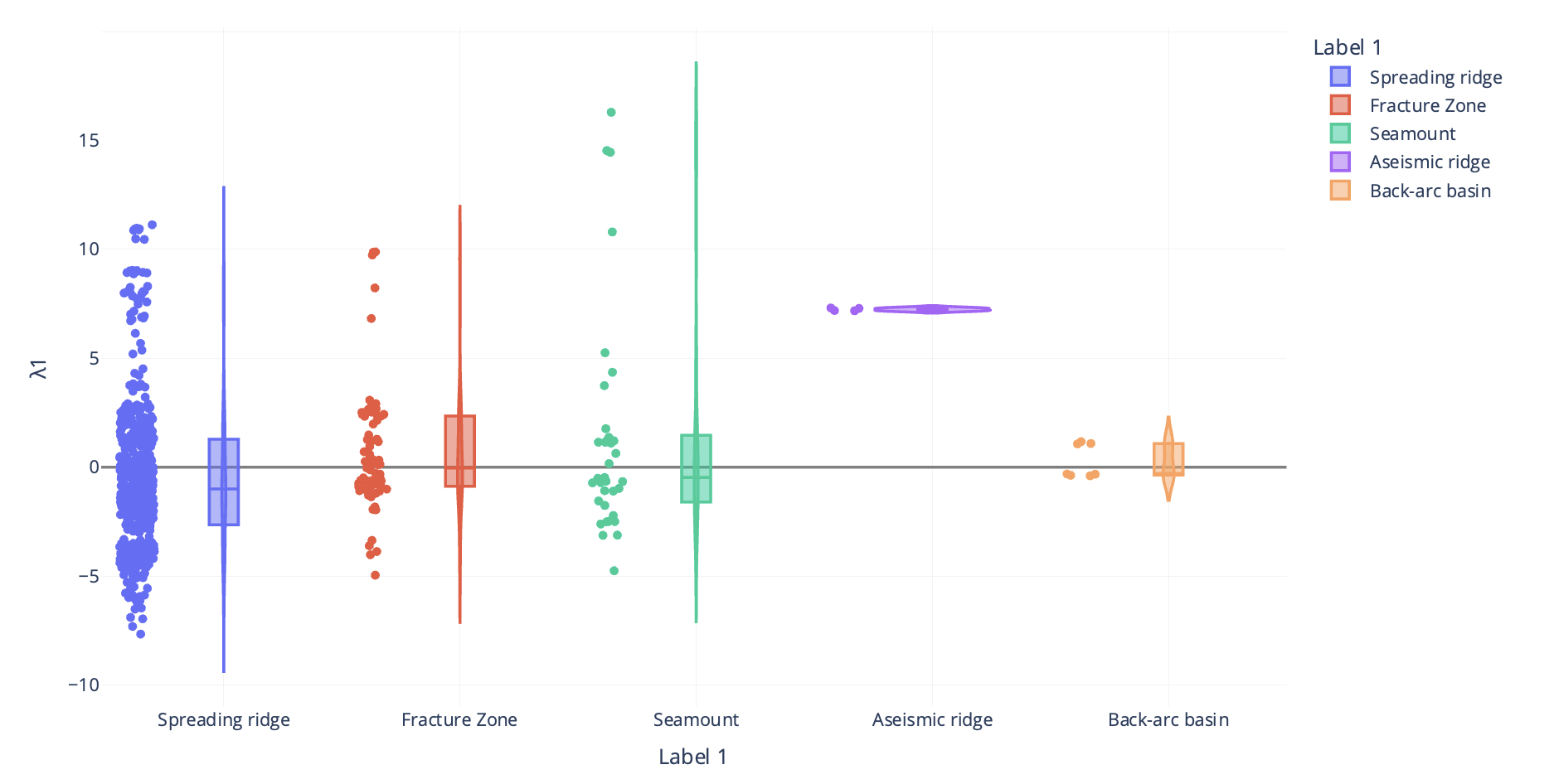}
\caption{Figure 21: Example violin plot taken from the REEkit application. The colours on the plot indicate the geological region of each sample (which is also shown on the X axis). The Y-axis represents lambda 1 (heavy or light REE enrichment).}
\end{figure}

\subsection{Effective combinations and standalone visualisations}

Much of the insight gained from discussions with participants related to how the visualisations and the REEkit platform would fit into their existing workflow for understanding REE data in their line of work. The choice and preference of visualisation was dependent on participants specific data analysis goals and the nature of their data. Moreover, by experimenting with various combinations of visualisations, participants were able to enhance their understanding of geochemical data, leading to more advanced data insights.

\subsubsection{Best standalone visualisations}

The majority of participants (greater than 50\%) identified two specific lambda visualisations as highly valuable standalone methods for discerning patterns and data populations – the scatter plot matrix and the 3D scatter plot.

\paragraph{Scatter plot matrix} The scatter plot matrix was highlighted as the top choice for a standalone visualisation method due to its ability to offer a thorough overview of data distribution in 2D space. Participants appreciated it not only for its widespread adoption in the industry but also for its effectiveness in facilitating clear communication with others. Its prevalence in the field of geochemistry made it easy to grasp, allowing participants to clearly see it being incorporated it into their existing analysis workflow.

\paragraph{3D scatter plot} Despite its controversial nature, the 3D aspect of the plot served as a valuable tool for participants, enabling them to quickly pinpoint the distribution of data concerning multiple lambda values. This feature not only provided valuable context but also allowed participants to explore and manipulate their data before delving into deeper analysis using other methods. 

\subsubsection{Best combinations of visualisations}

During the study some useful combinations of visualisations were mentioned and observed. One effective pairing was the 3D scatter and the density contour plot. These two methods worked well together, giving a comprehensive view of data. Starting with 3D scatter, participants could see where different groups were located in three-dimensional space. Then, density contour plots refined this understanding, revealing detailed relationships in a two-dimensional format. Although density contour plots hid individual data points, it provided a clear picture of patterns, improving participants' insights into how different data sets were interconnected. This led many participants to express insights that had been overlooked when using each visualisation on its own.

Similarly, the pairing of scatter plot matrix and the density contour plot received interest from several participants. Scatter plot matrix, with its capacity to visualise connections between multiple lambdas, combined with the insights from density contour plots into more focused data distributions between two lambdas, proved to be a powerful analytical approach for geochemical data analysis. Participants valued the viewpoint this combination provided.

Another valuable combination was found with scatter plot matrix and violin plots. This pairing struck a balance between delving into complex relationships within the data and gaining insight into data distribution within specific data categories (e.g., mineralogy). Violin plots, in particular, added depth, enabling a more detailed analysis of data variation across different groups, complimenting the data presented in the scatter matrix. 

It’s also worth mentioning that several participants commented on the combined utility of the three different variants of the scatter visualisation method. The integration of scatter plot with scatter plot matrix and 3D scatter plot, was identified as a useful toolkit for in-depth geochemical analysis by the majority of participants. 

Additionally, while spider diagram plots were deemed useful on their own, their utility significantly increased when combined with lambda data in the form of the 3D scatter plot. This combination provided a more holistic view of the data, simplifying the comprehension of intricate data patterns using lambdas and providing more detail where needed using the patterns in the spider diagram. 

\subsection{Further commentary}

\subsubsection{Stakeholder communication}

Most participants mentioned that a large part of their job revolved around communicating results and other important information with stakeholders involved in the mineral exploration industry. Stakeholders mentioned include managers, academics, industry professionals, other geologists, data analysts and investors.

In the context of our study, certain visualisation methods were commonly regarded as more effective for facilitating information discovery, while others were deemed more suitable for conveying specific points to an audience. Participants in the study generally concurred that simple 2D visualisations proved to be the most effective when communicating with non-technical stakeholders. Scatter 2D, density contour and scatter plot matrix were identified among the most useful in this regard. 

\subsubsection{Mineralogy and reference materials}

Within the imported CSV data, REEkit users can input different information which allows for the examination of data from various categorical perspectives. In participant discussions, identifying crucial categories for gaining insights with lambda data and accompanying information was explored.

Many participants emphasised various visualisations utility when assessing grouping of data in relation to mineralogy, with several stating that “mineralogy is king” in the exploration field. Participants emphasised the importance of comprehending the intricate relationship between patterns and mineralogy, especially when transitioning from an exploration discovery to a viable and economically feasible mineral resource. Additionally, it is noteworthy that several participants shared their ideas about internal labels and categories they envisaged adding to their internal datasets to use with REEkit and its visualisations.

Several participants also emphasised the importance of having standardised materials to compare their data and samples to when using the tool/visualisations. Participants suggested several materials including data from well-known deposits and common geological references like mid-ocean ridge basalt (MORB) or average crustal composition. Having such standard materials would enable participants to make meaningful comparisons and enhance the utility of the tool in their workflow.

\subsubsection{Participant feedback and suggestions}

Suggestions were made by participants to enhance their experience and understanding of the various visualisations. It was recommended that clearer explanations of how to use each visualisation be presented, ensuring that users comprehend the presented information, considering that not all users have interacted with each of the visualisation types before.
 
Additionally, participants suggested the implementation of tooltips across the suite of visualisations to clarify clusters and patterns, reducing the complexity of interpreting the lambdas. Features such as additional labels or annotations for data points, facilitating side-by-side comparison of different datasets, were also suggested to enhance analysis. 

Furthermore, participants recommended further customisation options for density contouring, including varied contour lines or the use of heat mapping. Additionally, users expressed the desire to incorporate features such as 3D lassoing, allowing the selection and manipulation of data points within a 3D space, and the ability to transfer selected data points between visualisations. These suggestions are discussed in more detail in section 6.1.

\section{Discussion}

\subsection{Visualisations and research objectives}

Participant responses from contextual inquiry revealed that lambda data significantly improved geochemical data analysis for mineral exploration in several ways:

\begin{itemize}
\item \textbf{Identifying patterns in REE data}: The lambda technique simplifies REE patterns, making it easier to categorise data into meaningful groups and understand larger patterns or trends in the dataset. This clearly made analysing larger geochemical REE datasets much simpler for participants in this study.
\item \textbf{Grouping similar analyses}: Visualisation methods like scatter plots and density contour plots, when used with lambda data, helped identify spatial data groupings. This ability to group analyses using lambda data can reveal variation within specific sites or projects, aiding in the development of petrogenetic models and understanding the categorisation of economically significant REEs.
\item \textbf{Understanding geochemical processes}: Participants used lambda data alongside visualisations to comprehend how underlying geochemical processes influenced data distribution. This approach facilitated quantitative analysis and highlighted differences between patterns.
\end{itemize}

The versatility of lambdas makes them valuable, with potential applications in various domains, spanning both academic research and industrial practices.

The most impactful standalone visualisations were those that participants were accustomed to, presented data in a straightforward manner, and effectively highlighted distinctions or similarities between data groups or individual points. This is because, in the field of mineral exploration, participants highlighted that an effective visualisation should serve both as a tool for data analysis and a means of communicating findings to others.

Among the visualisation methods assessed in this study, the most effective ones were the scatter plot matrix and the 3D scatter plot. While existing tools like ioGAS have incorporated similar visualisations, they haven't integrated them with lambda data. These findings suggest the potential to explore and assess additional visualisation techniques that present information in a similar fashion.

In addition to assessing individual visualisations, this study investigated how various visualisation methods could complement each other within the same workflow, providing participants and users with deeper insights.

Participants discovered that integrating varied visualisations deepened their comprehension of geochemical data, leading to more sophisticated insights. Effective combinations were those that provided distinct viewpoints on the presented data. Several participants casually remarked that in the realm of mineral exploration, 'any visualisation is a good visualisation,' indicating that complementary visualisations often offer a unique and fresh perspective on the data. 

This study demonstrated the effectiveness of different types of scatter plots when combined with visualisations highlighting data distribution (e.g. violin plots) or obscuring specific elements (e.g., density contour plots). Participants particularly emphasised the effectiveness of visualisations like scatter matrix and 3D scatter plots in these cases. 

Moreover, participants also highlighted the significance of having control visualisations (like spider diagrams) present throughout the analysis process. Having a familiar reference source, especially when working with lambda data, was consistently emphasised as crucial by the majority of participants.

These findings highlight the importance of familiarity and clarity in determining the effectiveness of visualisation methods when leveraging lambda data in mineral exploration. It's essential to carefully consider these factors when employing lambda data in both academic research and industrial applications. These insights will guide the ongoing development of the REEkit tool.

\subsection{The Future of REEkit Software}

REEkit, along with its associated visualisations, fill the role as an initial, first-pass, tool for handling geochemical REE data. Unlike many existing solutions that require extensive time, money, and effort to learn and implement, REEkit offers a straightforward, browser-based interface. It simplifies the process of comparing and analysing data, making it accessible to industry professionals, academics, and other stakeholders. With REEkit, users can quickly gain valuable insights from their data without the burden of complex setup, making it a practical choice for handling geochemical REE data effectively.

All participants unanimously recognised the practical implications of the tool for the REE industry. Their feedback emphasised the significance of REEkit and its visualisations, describing them as a positive advancement in communication within the REE sector. Comments included phrases like "a good step forward for communication in the REE industry" and the acknowledgment that "the industry needs to evolve and make everything more visual and easier to interpret." This consensus marked REEkit as a promising next stage and a crucial step forward for the industry, highlighting its potential to enhance communication and interpretation processes significantly.

REEkit was initially developed to align with the visualisation assessment objectives of this research project, focusing on the evaluation of various data visualisations. In its current state, the tool requires updated to cater to industry needs and user preferences discovered in this study. This update is crucial before advancing to the next version of the software. Additionally, the commercial viability of the next version remains uncertain, underscoring the importance of refining REEkit to be more industry-oriented and user-friendly before considering any further development or commercialisation efforts.

\subsection{Limitations}

Whilst this study resulted in clear conclusions and findings, it's important to acknowledge the limitations of the method employed. The contextual inquiry conducted in this study involved observing participants interact with visualisations through Zoom. Although this approach effectively offered valuable insights into typical user interactions with data, it did not encompass all potential scenarios or contexts pertinent to the research topic (i.e., when participants use data they have existing familiarity with), resulting in certain gaps in understanding. Furthermore, participants in this study were presented with pre-prepared data, a variation from the natural scenario a user would encounter while using a tool like REEkit in real-world situations.

Additionally, like many qualitative data collection methods, there's a possibility of unintentional subjectivity or bias in the interpretations and observations made during participant interactions. These interpretations could have been influenced by the interviewer's input and prompts, potentially introducing subjective bias into the findings.

Lastly, time and resource constraints hindered the exploration of a broader spectrum of contexts, underscoring the challenges faced in conducting thorough contextual inquiries. These limitations are important considerations in understanding the scope and applicability of the study's outcomes.

\section{Conclusion}

This study builds upon the research conducted by Anenburg and O’Neill by implementing the lambda method in a practical context. It adhered to a well-established qualitative data collection approach to assess the efficacy of various information visualisation techniques in conveying meaningful insights from geochemical REE data. 

This study emphasised the demand for innovative reporting and analytical tools in the field of REE exploration, with lambdas demonstrating notable effectiveness. The study revealed that two critical factors for successful geochemical data visualisation in the mineral exploration industry are familiarity and clarity. This is because effective visualisations must serve not only analytical purposes but also effectively communicate complex information to a non-technical audience. In this study, the 3D scatter plot and scatter plot matrix emerged as the most effective standalone visualisations. These visualisations, already prevalent in the field of geochemistry, proved exceptionally valuable for extracting insights from lambda data, emphasising their utility and relevance in this context.

The study revealed that participants' choice of visualisation was influenced by personal choice in workflows and analytical objectives. Visualisations like contour plots and distribution visualisations (such as violin plots) complemented the detailed perspectives provided by certain scatter plots. These findings underscore the importance of providing users with a diverse set of complementary visualisation methods within any analytical tool.

\subsection{Future Work}

Continuing research in this field is imperative. As REEs become an increasingly sought-after resource, the volume of associated data will inevitably surge, requiring the establishment of more effective and documented data visualisation methods.

The development of the REEkit platform should continue by incorporating the insights gleaned from this research. These findings should serve as a guiding principle when updating the platform to cater to the needs of the industry. 

Furthermore, it is important to investigate more complex and nuanced visualisation techniques and engage in contextual inquiries in different environments or conduct diary studies involving the use of REEkit. Future studies should observe how participant responses and attitudes change when using a familiar dataset. In future research endeavours aimed at exploring additional visualisation techniques, the REEkit platform should be leveraged as a practical tool for implementation and testing.

Ongoing research in this area will provide a deeper insight into the practical applications of visualisations in mineral exploration. This understanding is essential for the effective utilisation of REEs and their associated data in real-world scenarios within the industry.


\section*{References}

Adams, R. (2022) ‘We need 10 new REE mines by 2030 to meet magnet rare earths demand’, 26 September. Available at: \url{https://stockhead.com.au/resources/we-need-10-new-ree-mines-by-2030-to-meet-magnet-rare-earths-demand/}.

Agnerian, H. and Roscoe, W.E. (2002) ‘The contour method of estimating mineral resources’, CIM Bulletin, 95, pp. 100–107.

Anenburg, M. (2019) ‘Geochemistry data of Nolans Bore obtained by LA-ICP-MS’. figshare. Available at: \url{https://doi.org/10.6084/M9.FIGSHARE.5602720.V1}.

Anenburg, M. (2020) ‘Rare earth mineral diversity controlled by REE pattern shapes’, Mineralogical Magazine, 84(5), pp. 629–639. Available at: \url{https://doi.org/10.1180/mgm.2020.70}.

Busstra, C.M., Hartog, R. and Van ’T Veer, P. (2005) ‘Teaching: the role of active manipulation of three-dimensional scatter plots in understanding the concept of confounding’, Epidemiologic Perspectives \& Innovations, 2(1), p. 6. Available at: \url{https://doi.org/10.1186/1742-5573-2-6}.

Carpendale, S. (2008) ‘Evaluating Information Visualizations’, in A. Kerren et al. (eds) Information Visualization. Berlin, Heidelberg: Springer Berlin Heidelberg (Lecture Notes in Computer Science), pp. 19–45. Available at: \url{https://doi.org/10.1007/978-3-540-70956-5_2}.

Carr, M.J. and Gazel, E. (2017) ‘Igpet software for modeling igneous processes: examples of application using the open educational version’, Mineralogy and Petrology, 111(2), pp. 283–289. Available at: \url{https://doi.org/10.1007/s00710-016-0473-z}.
‘Configuration in Python’ (2023). Available at: \url{https://plotly.com/python/configuration-options/}.

Coryell, C.D., Chase, J.W. and Winchester, J.W. (1963) ‘A procedure for geochemical interpretation of terrestrial rare-earth abundance patterns’, Journal of Geophysical Research, 68(2), pp. 559–566. Available at: \url{https://doi.org/10.1029/JZ068i002p00559}.

Dushyantha, N. et al. (2020) ‘The story of rare earth elements (REEs): Occurrences, global distribution, genesis, geology, mineralogy and global production’, Ore Geology Reviews, 122, p. 103521. Available at: \url{https://doi.org/10.1016/j.oregeorev.2020.103521}.

Elfil, M. and Negida, A. (2017) ‘Sampling methods in Clinical Research; an Educational Review.’, Emergency (Tehran, Iran), 5(1), p. e52.

Erban, V. et al. (2003) ‘Geochemical Data Toolkit (GCDkit): a Key for Magmatic Geochemists to the Treasury of Data Analysis, Statistics and Graphics in R’, Geolines, 16, pp. 25–26.

Farmer, M. (2017) Migmatite delineates zones of melt flux through the upper crust, Wongwibinda, NSW. Macquarie University.

Gielen, D. and Lyons, M. (2022) Critical materials for the energy transition: Rare earth elements. International Renewable Energy Agency.

Godwin, L., Valleau, N. and Mortimer, D. (2021) ‘The evolution of geoscientific software — The past, present, and future’, in Symposium on the Application of Geophysics to Engineering and Environmental Problems 2021. Symposium on the Application of Geophysics to Engineering and Environmental Problems 2021, Online Conference: Society of Exploration Geophysicists and Environment and Engineering Geophysical Society, pp. 135–136. Available at: https://doi.org/10.4133/sageep.33-070.

Hintze, J.L. and Nelson, R.D. (1998) ‘Violin Plots: A Box Plot-Density Trace Synergism’, The American Statistician, 52(2), pp. 181–184. Available at: \url{https://doi.org/10.1080/00031305.1998.10480559}.

Jackson, R.G. (2010) ‘Application of 3D geochemistry to mineral exploration’, Geochemistry: Exploration, Environment, Analysis, 10(2), pp. 143–156. Available at: \url{https://doi.org/10.1144/1467-7873/09-217}.

Kennedy, M.E. (1998) ‘Elements: traceTrace’, in Geochemistry. Dordrecht: Kluwer Academic Publishers (Encyclopedia of Earth Science), pp. 221–223. Available at: \url{https://doi.org/10.1007/1-4020-4496-8_111}.

Kerren, A. et al. (eds) (2008) Information visualization: human-centered issues and perspectives. Seminar on Information Visualization - Human Centered Issues in Visual Representation, Interaction, and Evaluation, Berlin Heidelberg: Springer (Lecture notes in computer science, Vol. 4950).

Ladner, S. (2016) Practical Ethnography: A Guide to Doing Ethnography in the Private Sector. 0 edn. Routledge. Available at: \url{https://doi.org/10.4324/9781315422251}.

McLennan, S.M. (1994) ‘Rare earth element geochemistry and the “tetrad” effect’, Geochimica et Cosmochimica Acta, 58(9), pp. 2025–2033. Available at: \url{https://doi.org/10.1016/0016-7037(94)90282-8}.

Nazemi, K. et al. (2015) ‘Web-based Evaluation of Information Visualization’, Procedia Manufacturing, 3, pp. 5527–5534. Available at: \url{https://doi.org/10.1016/j.promfg.2015.07.718}.

O’Neill, H.St.C. (2016) ‘The Smoothness and Shapes of Chondrite-normalized Rare Earth Element Patterns in Basalts’, Journal of Petrology, 57(8), pp. 1463–1508. Available at: \url{https://doi.org/10.1093/petrology/egw047}.

Petrelli, M. et al. (2005) ‘PetroGraph: A new software to visualize, model, and present geochemical data in igneous petrology: PETROGRAPH SOFTWARE’, Geochemistry, Geophysics, Geosystems, 6(7), p. n/a-n/a. Available at: \url{https://doi.org/10.1029/2005GC000932}.

Rock, N.M.S. (1987) ‘The need for standardization of normalized multi-element diagrams in geochemistry: a comment’, Geochemical Journal, 21, pp. 75–84.

Rollinson, H. and Pease, V. (2021) Using Geochemical Data: To Understand Geological Processes. 2nd edn. Cambridge University Press. Available at: \url{https://doi.org/10.1017/9781108777834}.

Saunders, B. et al. (2018) ‘Saturation in qualitative research: exploring its conceptualization and operationalization’, Quality \& Quantity, 52(4), pp. 1893–1907. Available at: \url{https://doi.org/10.1007/s11135-017-0574-8}.

Scott, E.R.D. and Krot, A.N. (2007) ‘Chondrites and Their Components’, in Treatise on Geochemistry. Elsevier, pp. 1–72. Available at: \url{https://doi.org/10.1016/B0-08-043751-6/01145-2}.

Steichen, B. and Fu, B. (2019) ‘Towards Adaptive Information Visualization - A Study of Information Visualization Aids and the Role of User Cognitive Style’, Frontiers in Artificial Intelligence, 2, p. 22. Available at: \url{https://doi.org/10.3389/frai.2019.00022}.

Wang, X. et al. (2021) ‘DeHumor: Visual Analytics for Decomposing Humor’. Available at: \url{https://doi.org/10.48550/ARXIV.2107.08356}.

Williams, M. et al. (2020) ‘pyrolite: Python for geochemistry’, Journal of Open Source Software, 5(50), p. 2314. Available at: \url{https://doi.org/10.21105/joss.02314}.

Yu, Q.-Y. et al. (2019) ‘GeoPyTool: A cross-platform software solution for common geological calculations and plots’, Geoscience Frontiers, 10(4), pp. 1437–1447. Available at: \url{https://doi.org/10.1016/j.gsf.2018.08.001}.

\end{document}